\documentclass[a4paper,11pt]{article}
\usepackage{amsfonts}
\usepackage{amsmath}
\usepackage{amssymb}
\usepackage{mathtools}
\usepackage{colortbl}
\usepackage{cite}
\usepackage{hyperref}
\numberwithin{equation}{section}
\usepackage[left=2cm,right=2cm,top=3.5cm,bottom=3.5cm]{geometry}

\allowdisplaybreaks[1]
\hypersetup{
	colorlinks=true,
	linkcolor=ceruleanblue,
	filecolor=ceruleanblue,      
	urlcolor=ceruleanblue,
	citecolor=ceruleanblue,
}
\definecolor{ceruleanblue}{rgb}{0.0, 0.2, 0.6}

\date{\today}

\begin{document}
	
	\begin{flushright} {\footnotesize YITP-25-30, IPMU25-0009} \end{flushright}

	\begin{center}
		\LARGE{\bf Spherical black hole perturbations in EFT of scalar-tensor gravity with timelike scalar profile}
		\\[1cm] 
		
		\large{Shinji Mukohyama$^{\,\rm a, \rm b}$, Kazufumi Takahashi$^{\,\rm a}$, Keitaro Tomikawa$^{\,\rm c}$, \\ and Vicharit Yingcharoenrat$^{\,\rm b, \rm d}$}
		\\[0.5cm]
		
		\small{
			\textit{$^{\rm a}$
				Center for Gravitational Physics and Quantum Information, Yukawa Institute for Theoretical Physics, 
				\\ Kyoto University, 606-8502, Kyoto, Japan}}
		\vspace{.2cm}
		
		\small{
			\textit{$^{\rm b}$
				Kavli Institute for the Physics and Mathematics of the Universe (WPI), The University of Tokyo Institutes for Advanced Study (UTIAS), The University of Tokyo, Kashiwa, Chiba 277-8583, Japan}}
		\vspace{.2cm}
		
		\small{
			\textit{$^{\rm c}$
                Institute of Fundamental Physics and Quantum Technology,
Department of Physics, School of Physical Science and Technology,
Ningbo University, Ningbo, Zhejiang 315211, China
                }}
        \vspace{.2cm}
            
        \small{
            \textit{$^{\rm d}$
                High Energy Physics Research Unit, Department of Physics, Faculty of Science, Chulalongkorn University, Pathumwan, Bangkok 10330, Thailand}}
        \vspace{.2cm}
        
	\end{center}
	
	\vspace{0.3cm} 
	
	\begin{abstract}\normalsize
    We study linear even-parity perturbations of static and spherically symmetric black holes with a timelike scalar profile by use of the effective field theory (EFT) approach. For illustrative purposes, we consider a simple subclass of the EFT that accommodates ghost condensate, namely the k-essence model along with the so-called scordatura term, and focus on the spherical (monopole) perturbations about an approximately stealth Schwarzschild solution. The scordatura effect is introduced to avoid the strong coupling problem that typically happens in the scalar sector around stealth solutions with a timelike scalar profile. We argue that the scalar perturbation is decoupled from the metric perturbations at the leading order in the scordatura effect under a particular gauge choice.
    We stress that this is an important step in understanding the dynamics of even-parity perturbations, paving the way towards deriving a set of master equations---the generalized Zerilli and the scalar-field equations---for generic multipoles.
	\end{abstract}
	
	\vspace{0.3cm} 
	
	\vspace{2cm}
	
	\newpage
	{
		\hypersetup{linkcolor=black}
		\tableofcontents
	}
	
	\flushbottom
	
	\vspace{1cm}
	
	
\section{Introduction}

The study of modified theories of gravity has become an increasingly important topic in the research of cosmology and black hole (BH)~\cite{Koyama:2015vza,Ferreira:2019xrr,Arai:2022ilw}.
These modifications are often introduced in order to provide a clue to the Universe's unsolved mysteries, such as dark energy, inflation, and dark matter, which cannot be fully explained by general relativity (GR).
Furthermore, investigating the effects of modified gravity could provide valuable insights into constructing a consistent theory of quantum gravity at short scales. With many experiments and observations, particularly from gravitational wave (GW) detectors, e.g., LIGO-Virgo-KAGRA collaboration~\cite{LIGOScientific:2016aoc,LIGOScientific:2018mvr,LIGOScientific:2019lzm,KAGRA:2023pio}, one hopes to gain a better understanding of GR within the scales of observations and their accuracy. In other words, any deviations from GR can in principle be constrained by experiments/observations, helping to test the validity of GR.

One of the simplest modified gravity theories is scalar-tensor theory, described by a single scalar field and a spacetime metric. 
It has been extensively studied in the literature in the context of cosmology and BHs. 
The most general class of scalar-tensor theories with covariant second-order Euler-Lagrange equations is called Horndeski theory~\cite{Horndeski:1974wa,Deffayet:2011gz,Kobayashi:2011nu}, and then some beyond Horndeski theories were found in \cite{Zumalacarregui:2013pma,Gleyzes:2014dya}.
An extension of the Horndeski theory, in which the Euler-Lagrange equations contain higher derivatives, is the so-called degenerate higher-order scalar-tensor (DHOST) theories~\cite{Langlois:2015cwa,Crisostomi:2016czh,BenAchour:2016fzp,Takahashi:2017pje,Langlois:2018jdg}.
(See also \cite{Langlois:2018dxi,Kobayashi:2019hrl} for comprehensive reviews and references therein.)
Note that, in DHOST theories, one avoids the Ostrogradsky ghosts~\cite{Woodard:2015zca} by imposing the degeneracy conditions~\cite{Motohashi:2014opa,Langlois:2015cwa,Motohashi:2016ftl,Klein:2016aiq} on the higher-derivative terms.
Furthermore, by requiring the degeneracy condition only under the unitary gauge, it can be generalized to the U-DHOST theories~\cite{DeFelice:2018ewo,DeFelice:2021hps,DeFelice:2022xvq}.
Moreover, performing a higher-order generalization of the invertible disformal transformations~\cite{Takahashi:2021ttd,Takahashi:2023vva} on Horndeski theories or U-DHOST theories gives rise to the generalized disformal Horndeski~\cite{Takahashi:2022mew} and generalized disformal unitary-degenerate theories~\cite{Takahashi:2023jro}, for which the conditions for consistent matter coupling were discussed in \cite{Takahashi:2022mew,Naruko:2022vuh,Takahashi:2022ctx,Ikeda:2023ntu,Takahashi:2023vva}.

All these covariant scalar-tensor theories in principle should be put in a single framework, called effective field theory (EFT). 
The formulation of the EFT relies on the symmetry breaking pattern and the background of the field contents.
An EFT of scalar-tensor theories on Minkowski/de Sitter background was constructed in \cite{Arkani-Hamed:2003pdi,Arkani-Hamed:2003juy} (called the EFT of ghost condensation) with the assumption that the timelike derivative of a scalar field spontaneously breaks the temporal diffeomorphism invariance, leaving the action invariant under the spatial diffeomorphism. 
Such an EFT was extended to a cosmological background, described by the Friedmann-Lema{\^i}tre-Robertson-Walker (FLRW) metric. 
As a result, one obtains the so-called EFT of inflation/dark energy~\cite{Cheung:2007st,Gubitosi:2012hu}, which plays a crucial role in describing the dynamics of perturbations on cosmological scales.
It is of course possible that this EFT breaks down at the scales relevant to BH physics.
In this case, one has to prepare another EFT to describe BHs, and this EFT is independent of the EFT of cosmological perturbations unless one knows an ultraviolet (UV) completion that accommodates the two EFTs.
However, it would be intriguing to develop an EFT framework that applies to physics both in the regimes of cosmology and BHs.
This led the authors of \cite{Mukohyama:2022enj} to construct an EFT of perturbations about an arbitrary background with a timelike scalar profile.\footnote{See \cite{Franciolini:2018uyq} for the formulation of the EFT of BH perturbations with a spacelike scalar profile (see also \cite{Hui:2021cpm} for an extension of the EFT to a slowly rotating BH, and \cite{Kuntz:2020yow} for further discussion on the EFT of \cite{Franciolini:2018uyq} and its application to extreme mass ratio inspiral systems).
Moreover, the EFT of BH perturbations for vector-tensor gravity has also been formulated recently in~\cite{Aoki:2023bmz}, extending the EFT on a cosmological background~\cite{Aoki:2021wew}.}
The shift-symmetric version of such an EFT was formulated in \cite{Khoury:2022zor}.
Using such an EFT, it is possible to extract information about dark energy from BH physics in principle, particularly from merger events.\footnote{In general, one may employ different frames depending on the regime of interest (e.g., cosmology or BHs), and hence one has to perform a frame transformation to compare the results obtained in different frames. The authors of \cite{Mukohyama:2024pqe} studied the conformal/disformal transformation of the EFT to address this issue.}
In \cite{Mukohyama:2022skk}, the so-called generalized Regge-Wheeler equation was derived, which is a master equation for describing dynamics of odd-parity perturbations about a static and spherically symmetric BH. Subsequently, the quasinormal mode spectrum for the odd sector was first obtained in \cite{Mukohyama:2023xyf} in the context of the EFT (see \cite{Konoplya:2023ppx} for investigation of the spectrum of higher overtones), and then graybody factors were also studied in \cite{Konoplya:2023ppx,Oshita:2024fzf}.
Moreover, it was shown in \cite{Barura:2024uog} that 
the so-called tidal Love numbers are in general non-vanishing in the EFT, which could be a smoking gun of deviations from GR in four dimensions.
However, the analysis of even-parity perturbations based on the EFT is still lacking.
The aim of the present paper is to make a first step to fill this gap.

The dynamics of both odd- and even-parity perturbations around the so-called stealth background~\cite{Mukohyama:2005rw,Motohashi:2018wdq,Takahashi:2020hso,Kobayashi:2025evr}, which is an exact solution of the spacetime metric having the same form as in GR, within the class of shift-symmetric Horndeski and (D)HOST theories has been studied in detail in the literature, e.g.,~\cite{Ogawa:2015pea,Takahashi:2016dnv,Babichev:2018uiw,Takahashi:2019oxz,deRham:2019gha,Khoury:2020aya,Tomikawa:2021pca,Langlois:2021aji,Takahashi:2021bml,Nakashi:2022wdg,Langlois:2022ulw,Nakashi:2023vul}.
It was found that the perturbations about stealth solutions in Horndeski and DHOST theories are strongly coupled in the asymptotic Minkowski/de Sitter region due to the vanishing of sound speed~\cite{deRham:2019gha,Khoury:2020aya,Takahashi:2021bml}, as expected from the results of \cite{Arkani-Hamed:2003pdi} based on the EFT of ghost condensation.
A possible way to avoid this issue is to take into account a small detuning of (one of) the degeneracy conditions (i.e., the scordatura term), which introduces a higher derivative term, corresponding to a $k^4$~term in the dispersion relation~\cite{Motohashi:2019sen}.
(See \cite{Gorji:2020bfl,Gorji:2021isn} for the study of scordatura term in the context of cosmology.)
Note that the scordatura mechanism is built-in in the EFT of ghost condensation~\cite{Arkani-Hamed:2003pdi,Mukohyama:2005rw} and in the U-DHOST theories~\cite{DeFelice:2022xvq}.
Also, it was shown in \cite{Mukohyama:2005rw,DeFelice:2022qaz} that the scordatura term leads to a small deviation from the exactly stealth configuration.
In the present paper, we study the dynamics of even-parity perturbations in the EFT, taking into account the scordatura term to avoid the strong coupling problem.
For simplicity, we will focus on the spherical (monopole) perturbations and investigate the structure of the equations of motion (EOMs) for the perturbations.
We argue that the scalar perturbation is decoupled from the metric perturbations at the leading order in the scordatura effect under a particular gauge choice.
Our study in this paper is an important step towards deriving a set of master equations---the generalized Zerilli and the scalar-field equations---for generic multipoles without any strong coupling problem, offering crucial insights into the dynamics and phenomenology of even-parity perturbations in the presence of the scordatura term.

The rest of the paper is organized as follows.
In Section~\ref{sec:setup}, we briefly review the EFT of perturbations on an arbitrary background with a timelike scalar profile and describe the setup of our model.
In Section~\ref{sec:even_sector}, we discuss even-parity perturbations about the approximately stealth Schwarzschild solution, with a particular focus on spherical (monopole) perturbations.
Finally, we draw our conclusions in Section~\ref{sec:conclusion}.

\section{Setup}\label{sec:setup}  
\subsection{EFT of scalar-tensor gravity with timelike scalar profile}
In this Section, we explain a general approach to study the dynamics of perturbations on an arbitrary background with a timelike scalar field in the effective field theory (EFT) framework~\cite{Mukohyama:2022enj}. 
One of the advantages of the EFT approach is that it unifies scalar-tensor theories in a way that their effects on the dynamics of perturbations are represented by a finite number of coefficients. 

Let us consider generic scalar-tensor theories described by the metric~$g_{\mu\nu}$ and the scalar field~$\Phi$, where $\partial_\mu\Phi$ is assumed to be timelike.
In the unitary gauge where $\delta \Phi \equiv \Phi - \bar{\Phi}(\tau) = 0$, with $\tau$ being the time coordinate in terms of which the scalar field is spatially uniform, the EFT action can be written as~\cite{Mukohyama:2022enj,Mukohyama:2022skk}
\begin{align}\label{eq:action_EFT}
    S=S_{(0)}+S_{(2)}\;,
\end{align}
with 
\begin{align}
    S_{(0)} &= \int {\rm d}^4 x \sqrt{-g} \bigg[\frac{M_\star^2}{2} f(x) R - \Lambda(x) - c(x)g^{\tau\tau} - \tilde{\beta}(x) K - \tilde{\alpha}^\mu_\nu(x) K^\nu_\mu - \zeta(x) n^\mu\partial_\mu g^{\tau\tau}\bigg]  \;, \label{eq:S0} \\
    S_{(2)} &= \int {\rm d}^4 x \sqrt{-g}\bigg[\frac{1}{2} m_2^4(x) (\delta g^{\tau\tau})^2 + \frac{1}{2} \tilde{M}_1^3(x) \delta g^{\tau\tau} \delta K + \frac{1}{2} M_2^2(x) \delta K^2 + \frac{1}{2} M_3^2(x) \delta K^\mu_\nu \delta K^\nu_\mu \nonumber \\
	&\qquad\qquad\qquad~~+ \frac{1}{2}\mu_1^2(x) \delta g^{\tau\tau} \delta {}^{(3)}\!R + \frac{1}{2} \lambda_1(x)^\mu_\nu \delta g^{\tau\tau} \delta K^\nu_\mu + \frac{1}{2}{\mathcal M}_1^2(x)(\bar{n}^\mu\partial_\mu\delta g^{\tau\tau})^2 \nonumber \\
	&\qquad\qquad\qquad~~+\frac{1}{2}{\mathcal M}_2^2(x)\delta K(\bar{n}^\mu\partial_\mu\delta g^{\tau\tau})+\frac{1}{2}{\mathcal M}_3^2(x)\bar{h}^{\mu\nu}\partial_\mu\delta g^{\tau\tau}\partial_\nu\delta g^{\tau\tau}\bigg]\;,
\end{align}
where 
a bar refers to the background value and $R$ is the 4d Ricci scalar. 
Also, $n_\mu\equiv -\delta_\mu^\tau/\sqrt{-g^{\tau\tau}}$ denotes the unit vector normal to a constant-$\tau$ hypersurface, $h_{\mu\nu}\equiv g_{\mu\nu}+n_\mu n_\nu$ is the induced metric, $K_{\mu\nu}\equiv h_\mu^\alpha\nabla_\alpha n_\nu$ is the extrinsic curvature, and ${}^{(3)}\!R$ is the 3d Ricci scalar.
Moreover, we have defined $\delta g^{\tau\tau} \equiv g^{\tau\tau} - \bar{g}^{\tau\tau}(x)$, $\delta K^\mu_\nu \equiv K^\mu_\nu - \bar{K}^\mu_\nu(x)$, and $\delta K\equiv \delta K^\mu_\mu$.
We note that the background values of $g^{\tau\tau}$ and $K^\mu_\nu$ as well as the EFT coefficients are functions of both temporal and spatial coordinates (collectively denoted by $x$) in general.
Such an EFT Lagrangian can be systematically obtained by Taylor-expanding a Lagrangian of the form~$L(g^{\tau\tau},K^\mu_\nu,{}^{(3)}\!R^\mu_\nu,\nabla_\mu,\tau)$, with ${}^{(3)}\!R_{\mu\nu}$ being the 3d Ricci tensor, that respects the 3d diffeomorphism invariance, i.e., the residual symmetry of our EFT in the unitary gauge.
Note that we have omitted operators such as $\delta K \delta{}^{(3)}\!R$, $\delta K^\mu_\nu \delta{}^{(3)}\!R^\nu_\mu$, and $\bar{K}^\mu_\nu\delta K^\nu_\alpha\delta K^\alpha_\mu$ in $S_{(2)}$ for simplicity.
As clarified in \cite{Mukohyama:2022skk}, Eq.~\eqref{eq:action_EFT} is a minimal EFT action that accommodates shift-symmetric quadratic HOST theories.
See \cite{Mukohyama:2022enj,Mukohyama:2022skk} for the dictionary between covariant theories (e.g., Horndeski and shift-symmetric HOST theories) and the set of EFT parameters.

As mentioned above, the residual symmetry in the unitary gauge is the 3d diffeomorphism invariance, and hence the action~\eqref{eq:action_EFT} should respect this symmetry.
However, the fact that in this setup the background values of the EFT building blocks are functions of both $\tau$ and $\vec{x}$ implies that each term of the EFT written in the unitary gauge breaks the 3d diffeomorphism invariance.
Therefore, as explained in \cite{Mukohyama:2022enj,Mukohyama:2022skk}, in order for the EFT action~\eqref{eq:action_EFT} to be invariant under the 3d diffeomorphism, one necessarily imposes a set of consistency relations among the EFT parameters that follows from applying the chain rule associated with the spatial derivative acting on each Taylor coefficient.
Note that the consistency relations are automatically satisfied for covariant theories such as Horndeski and HOST theories.

In Eq.~\eqref{eq:action_EFT}, the action~$S_{(0)}$ contains the tadpole terms, which describes the background dynamics. 
In fact, the variation of $S_{(0)}$ with respect to the metric yields the background equations~\cite{Mukohyama:2022enj}, and they put a set of constraints among the EFT parameters~$f$, $\Lambda$, $c$, $\tilde{\beta}$, $\tilde{\alpha}^\mu_\nu$, and $\zeta$ once the background metric functions are given.

On the other hand, the action~$S_2$ contains terms that are quadratic in perturbations.
Let us comment on the first few terms.
The $m_2^4$~operator is associated with the k-essence model. 
The operator~$\tilde{M}_1^3 \delta g^{\tau\tau} \delta K$ is associated with, for instance, the cubic Galileon theory.
It leads to several interesting phenomena of perturbations, e.g., the instabilities of dark energy induced by GWs~\cite{Creminelli:2019kjy} in the context of the EFT of dark energy.  
The $M_2^2$~operator is associated with the operator~$(\Box \Phi)^2$ in a covariant theory, which is nothing but the scordatura term~\cite{Motohashi:2019ymr}.
The operator~$M_3^2 \delta K^\mu_\nu \delta K^\nu_\mu$ typically affects the speed of tensor perturbations.
For example, on a static and spherically symmetric background, this term modifies the speed of GWs in the odd-parity sector~\cite{Mukohyama:2022skk,Mukohyama:2023xyf}.
One expects that this $M_3^2$~term would also affect the speed of GWs in the even sector. 

Let us explain a general strategy to study the dynamics of linear perturbations about a given background using the EFT~(\ref{eq:action_EFT}). 
First, we fix both the backgrounds of the metric and the scalar field, which results in fixing (a part of) the tadpole coefficients through the background equations.
Then, one introduces $\delta g_{\mu\nu} = g_{\mu\nu} - \bar{g}_{\mu\nu}$ and expands the EFT action up to the quadratic order in the perturbations, from which the EOMs for the perturbations can be obtained.
Notice that, in the unitary gauge, the scalar fluctuation is set to zero ($\delta\Phi = 0$), and all the perturbations are inside the metric sector.
For example, in \cite{Takahashi:2021bml}, the analysis of even-parity perturbations about stealth Schwarzschild(-de Sitter) solutions in DHOST theories was carried out in the unitary gauge.
On the other hand, it was shown in \cite{deRham:2019gha} that, within a particular subclass of DHOST theories, there exists a (non-unitary) gauge where the scalar perturbation is decoupled from the metric perturbations.
In Section~\ref{sec:even_sector}, we will work in such a gauge and see that the decoupling occurs at the leading order in the scordatura effect.
For this purpose, it would be useful to recover the scalar fluctuation in the EFT action~\eqref{eq:action_EFT}, which can be achieved by performing the Stueckelberg trick associated with the non-linearly realized time diffeomorphism invariance: $\tau \rightarrow \tau + \pi(\tau,\vec{x})$ with $\pi(\tau, \vec{x})$ being the associated Nambu-Goldstone boson.
See \cite{Mukohyama:2022enj} for the Stueckelberg procedure on a generic background.\footnote{See also \cite{Cusin:2017mzw} for a similar trick valid on an FLRW background up to the second order in $\pi$.}
This is a general procedure in order to recover the full 4d diffeomorphism invariance of the EFT.
Indeed, the Nambu-Goldstone boson~$\pi(\tau, \vec{x})$ corresponds to the scalar fluctuation~$\delta \Phi$.

\subsection{Minimal EFT with scordatura term}\label{sec:k-essence_scor}

In the rest of this paper, for illustrative purposes, we focus on a simple subclass of the EFT that accommodates the k-essence model with the scordatura term, i.e.~ghost condensation~\cite{Arkani-Hamed:2003pdi,Arkani-Hamed:2003juy}.
We stress that this minimal EFT allows us to capture the essential features of the EFT around a stealth or approximately stealth background in general, and it would be straightforward to generalize the discussion in the present paper to a more general class of the EFT.

The action of the k-essence model with the scordatura term is given by
\begin{align}\label{eq:action_scorda}
\tilde{S} = \int \mathrm{d}^4x \sqrt{-g} ~\bigg[\frac{M_{\rm Pl}^2}{2} R + F_0(X) + \frac{\alpha_{\rm L}  (\Box \Phi)^2}{\Lambda_{\rm s}^2} \bigg] \;,
\end{align}
where $M_{\rm Pl}$ denotes the Planck mass, $F_0(X)$ is an arbitrary function of $X\equiv g^{\mu\nu}\partial_\mu\Phi\partial_\nu\Phi$, $\Lambda_{\rm s}$ is an energy scale associated with the scordatura term, and $\alpha_{\rm L}$ is a dimensionless parameter of order unity. 
Clearly, the action presented above is invariant under $\Phi \rightarrow \Phi + const.$ and $\Phi \rightarrow -\Phi$.
Thanks to the dictionary obtained in \cite{Mukohyama:2022enj,Mukohyama:2022skk}, we find that the action~\eqref{eq:action_scorda} is embedded in our EFT as
\begin{equation}\label{eq:dic_scordatura}
    \begin{aligned}
    &M_\star^2 f = M^2_{\rm Pl} \;, \quad
    \Lambda = -\bar{F}_0 + \bar{X} \bar{F}_{0X} - \frac{2 \alpha_{\rm L} }{\Lambda_{\rm s}^2}\bar{X} \bar{K}^2 \;, \quad
    c = \bar{X}\bar{F}_{0X} - \frac{\alpha_{\rm L}}{\Lambda_{\rm s}^2} \bar{X} \bar{K}^2 \;, \quad
    \tilde{\beta} = \frac{2\alpha_{\rm L}}{\Lambda_{\rm s}^2}\bar{X}\bar{K} \;,\\
    &\tilde{\alpha}^\mu_\nu=0\;, \quad
    \zeta=-\frac{\alpha_{\rm L}}{\Lambda_{\rm s}^2}\bar{X}\bar{K}\;, \quad
    m_2^4 = \bar{X}^2 \bar{F}_{0XX} \;, \quad
    \tilde{M}_1^3 = \frac{4\alpha_{\rm L}}{\Lambda_{\rm s}^2}\bar{X} \bar{K} \;, \quad M_2^2 = - \frac{2 \alpha_{\rm L}}{\Lambda_{\rm s}^2}\bar{X} \;, \\
    &M_3^2=0\;, \quad
    \mu_1^2=0\;, \quad
    \lambda_1{}^\mu_\nu=0\;, \quad
    {\cal M}_1^2=-\frac{\alpha_{\rm L}}{2\Lambda_{\rm s}^2}\bar{X}\;, \quad
    {\cal M}_2^2=\frac{2\alpha_{\rm L}}{\Lambda_{\rm s}^2}\bar{X}\;, \quad
    {\cal M}_3^2=0\;,
\end{aligned}
\end{equation}
where a subscript~$X$ refers to the derivative with respect to $X$, e.g., $F_{0X}={\rm d} F_0/{\rm d} X$.
Notice that 
we have assumed a background where $\bar{\Phi}(\tau) \propto \tau$ and $\bar{g}^{\tau\tau} = -1$, so that $\bar{X}$ is a constant.
It should also be noted that the k-essence and the scordatura terms in Eq.~(\ref{eq:action_scorda}) do not modify the speed of tensor perturbations, as is clear from the fact that $M_3^2=0$ [recall that $M_3^2$ is the coefficient of $\delta K^\mu_\nu \delta K^\nu_\mu$ in the EFT action].
In what follows, we focus on the minimal subclass of the EFT characterized by the EFT coefficients in Eq.~\eqref{eq:dic_scordatura}.

Let us comment on the effect of the scordatura term, which is controlled by $\alpha_{\rm L}/\Lambda_{\rm s}^2$.
The presence of the scordatura term implies that the existence of apparent Ostrogradsky ghost(s) in general but that it is at least as heavy as the energy scale~$\Lambda_{\rm s}$ that plays the role of the cutoff of our EFT. This means that the apparent Ostrogradsky ghosts are outside the regime of validity of the EFT and thus should be integrated out. 
Then, within the validity of the EFT, one can treat the effect of the scordatura term perturbatively.
(Therefore, even though the dimensionless parameter~$\alpha_{\rm L}$ itself can be of order unity, we will use it as a bookkeeping parameter for the perturbative expansion with respect to the scordatura effect.)
Nevertheless, a non-trivial thing happens in the EOM for the scalar perturbation:
As we will explicitly see in Section~\ref{sec:even_sector}, the scordatura term provides 
a fourth spatial-derivative term in the EOM of $\delta\Phi$.
Note that, in the asymptotic flat (or de Sitter) region of the stealth background, the second spatial-derivative term is typically suppressed by the Planck scale~\cite{Motohashi:2019ymr}, which is the origin of the strong coupling problem, and therefore the fourth spatial-derivative term is leading.
Actually, this is why the scordatura term can cure the strong coupling problem for the stealth background.
It should also be noted that, as we will see shortly in the next Subsection, the scordatura term leads to a deviation from the exactly stealth configuration~\cite{Mukohyama:2005rw,DeFelice:2022qaz}.

\subsection{Approximately stealth Schwarzschild background}\label{sec:BG}

In this Subsection, we briefly discuss the background configuration of both the scalar field~$\Phi$ and the metric~$g_{\mu\nu}$ based on our minimal EFT, i.e., the action~\eqref{eq:action_EFT} with the EFT coefficients given by Eq.~\eqref{eq:dic_scordatura}.

First, in the absence of the scordatura term ($\alpha_{\rm L} = 0$), let us discuss the stealth Schwarzschild solution, for which the metric and the scalar field are given by
\begin{equation}\label{eq:bg_sol}
\begin{aligned}
    &\bar{g}^{(0)}_{\mu\nu} {\rm d} x^\mu {\rm d} x^\nu = -A(r){\rm d} t^2 + \frac{{\rm d} r^2}{A(r)} + r^2 \gamma_{ab} {\rm d} x^a {\rm d} x^b \;, \qquad A(r) = 1 - \frac{r_{\rm s}}{r} \;, \\
    &\bar{\Phi}^{(0)}(t, r) = q t + \psi(r) \;, \qquad \bar{X}^{(0)} = -q^2 \;,
\end{aligned}
\end{equation}
where a bar denotes the background values, 
$\gamma_{ab}$ is the metric on a unit two-sphere with $a,b\in\{\theta, \phi\}$, $r_{\rm s}$ is the Schwarzschild radius, and $q$ is a non-vanishing constant.
Also, we have put the superscript~$(0)$ to imply that the result is of zeroth order in $\alpha_{\rm L}$.
Provided that $q^2>0$, the scalar field is timelike everywhere, i.e., $X < 0$, and the radial function~$\psi(r)$ can be explicitly written as~\cite{Mukohyama:2005rw}
\begin{align}
    \psi(r) = q \int \frac{\sqrt{1-A}}{A} {\rm d} r \;.
\end{align}
The existence conditions for the stealth Schwarzschild solution were obtained in \cite{Mukohyama:2022skk}, which in the present case read $\Lambda=c=0$, i.e.,
\begin{align}\label{eq:con_F0}
    \bar{F}^{(0)}_0\equiv F_0(-q^2) = 0 \;, \qquad
    \bar{F}^{(0)}_{0X}\equiv F_{0X}(-q^2) = 0 \;.
\end{align}
It is important to note that, in the absence of $\alpha_{\rm L}$, the sound speed of the scalar fluctuation is typically proportional to the parameter~$c$, and hence the vanishing of $c$ would lead to the strong coupling problem. In addition to the conditions (\ref{eq:con_F0}), we require $\bar{F}^{(0)}_{0XX}\equiv F_{0XX}(-q^2) > 0$ so that the fluctuation of the scalar field around the background has positive kinetic energy.\footnote{In our setup these conditions exclude a special class of scalar-tensor theories where the scalar field does not propagate, which is known as the (extended) cuscuton~\cite{Afshordi:2006ad,Iyonaga:2018vnu}.}

For later convenience, we express the background solutions given in Eq.~(\ref{eq:bg_sol}) in terms of the Painlev\'e-Gullstrand (PG) coordinates~$(\tau, r, \theta, \phi)$ as
\begin{align}\label{eq:bg_PG}
    \bar{g}^{(0)}_{\mu\nu} {\rm d} x^\mu {\rm d} x^\nu = -A {\rm d}\tau^2 + 2 \sqrt{1-A}\, {\rm d}\tau {\rm d} r + {\rm d} r^2 + r^2 \gamma_{ab} {\rm d} x^a {\rm d} x^b \;, \qquad \bar{\Phi}^{(0)}(\tau) = q \tau \;,
\end{align}
where the time coordinate~$\tau$ is related to $t$ and $r$ via
\begin{align}
    {\rm d}\tau = {\rm d} t + \frac{\sqrt{1 - A}}{A} {\rm d} r \;.
\end{align}
It is important to note that the background scalar field is a function only of $\tau$, and therefore $\tau$ here can be identified with that we have used in the construction of the EFT in the previous Subsections.
Moreover, similarly to the coordinate system~$(t, r, \theta, \phi)$, the background metric explicitly reduces to the Minkowski metric for large $r$.

Let us also introduce the Lema{\^i}tre coordinates~$(\tau, \rho, \theta, \phi)$, where the background configuration takes the form
\begin{align}\label{eq:bg_Lemai}
    \bar{g}^{(0)}_{\mu\nu} {\rm d} x^\mu {\rm d} x^\nu = -{\rm d}\tau^2 + (1-A){\rm d} \rho^2 + r^2 \gamma_{ab} {\rm d} x^a {\rm d} x^b \;, \qquad \bar{\Phi}^{(0)}(\tau) = q \tau \;.
\end{align}
Here, the coordinate~$\rho$ is related to $\tau$ and $r$ as 
\begin{align}\label{eq:Lemaitre_trans}
    {\rm d}\rho = {\rm d}\tau + \frac{{\rm d} r}{\sqrt{1-A}} \;.
\end{align}
The coordinate~$\rho$ is spacelike everywhere outside the horizon: $\bar{g}^{(0)\mu\nu}\partial_\mu\rho\partial_\nu\rho=(1-A)^{-1}>0$.
Equation~\eqref{eq:Lemaitre_trans} tells us that the areal radius~$r$ is a function of $\rho - \tau$.
We note that, in the asymptotic limit~$r \rightarrow \infty$, the $(\rho\rho)$-component of the metric~(\ref{eq:bg_Lemai}) vanishes, which implies that the proper distance on a constant-$\tau$ hypersurface decreases to zero.
Therefore, unlike the PG coordinates, the Lema{\^i}tre coordinates are not suitable when discussing the asymptotic behaviors of fluctuations.
However, as we will see in Section~\ref{sec:even_sector}, the Lema{\^i}tre coordinates are useful to understand the importance of the scordatura term.

Let us now consider the background solutions in the presence of the scordatura term.
As was pointed out in \cite{Mukohyama:2005rw,DeFelice:2022qaz}, when $\alpha_{\rm L}\ne 0$, the stealth configuration~\eqref{eq:bg_sol} ceases to be an exact solution, and the solution is subjected to corrections due to $\alpha_{\rm L}$.
The spherically symmetric solution up to the first order in $\alpha_{\rm L}$ can be written in the form~\cite{Mukohyama:2005rw,DeFelice:2022qaz}
    \begin{equation} \label{eq:corrected_BG_PG}
    \begin{split}
    &\bar{g}_{\tau\tau}= -A(r)[1-\alpha_{\rm L}h_0(\tau, r)] \;, \qquad
    \bar{g}_{\tau r}=\sqrt{1-A(r)}\,[1+\alpha_{\rm L}h_1(\tau, r)] \;, \\
    &\bar{g}_{rr}=1+\alpha_{\rm L}h_2(\tau, r) \;, \qquad
    \bar{g}_{\tau a}=\bar{g}_{r a}=0\;, \qquad
    \bar{g}_{ab}=r^2\gamma_{ab}\;,
    \end{split}
    \end{equation}
where $h_0$, $h_1$, and $h_2$ quantify the deviation from the (exactly) stealth Schwarzschild background expressed in terms of the PG coordinates.
The components~$\bar{g}_{\tau a}$ and $\bar{g}_{r a}$ do not receive corrections due to the spherical symmetry.
Note that the corrections to the background scalar field~$\bar{\Phi}(\tau)$ can always be set to zero, which amounts to redefining the time coordinate.
The corrections to the background metric can be treated perturbatively if $|\alpha_{\rm L}h_i|\ll 1$ for $i=0,1,2$.\footnote{In the Lema{\^i}tre coordinates, one would define the corrections due to $\alpha_{\rm L}$ as $\bar{g}_{\tau\tau}=-(1-\alpha_{\rm L}\tilde{h}_0)$, $\bar{g}_{\tau\rho}=\alpha_{\rm L}\tilde{h}_1$, and $\bar{g}_{\rho\rho}=(1-A)(1+\alpha_{\rm L}\tilde{h}_2)$.
In this case, the criterion for the smallness of the corrections is non-trivial as $\alpha_{\rm L}\tilde{h}_1$ is the leading contribution to $\bar{g}_{\tau \rho}$.}
For the explicit expression of the EOMs for $h_0$, $h_1$, and $h_2$, see \cite{DeFelice:2022qaz}.\footnote{\label{footnote:BGcorrections}
Precisely speaking, the EOMs in \cite{DeFelice:2022qaz} were derived for a subclass of U-DHOST theories where the scordatura effect is built-in, whose action has the form
\begin{align*}
\tilde{S}_\text{U-DHOST} = \int \mathrm{d}^4x \sqrt{-g} ~\bigg[\frac{M_{\rm Pl}^2}{2} R + F_0(X) + 
A_2(X) \left(\Box \Phi-\frac{1}{2X}\nabla^\mu\phi\nabla_\mu X\right)^2 \bigg] \;,
\end{align*}
with $A_2$ being a function of $X$.
If we identify $A_2$ as $\alpha_{\rm L}/\Lambda_{\rm s}^2$, the difference from our model~\eqref{eq:action_scorda} is the last term inside the square brackets.
Nevertheless, we have confirmed that the EOMs of the background corrections derived from our model precisely coincides with those in \cite{DeFelice:2022qaz}.
Note that, in more generic HOST theories, the effect of the additional interactions is suppressed, and the solution is essentially described by the one for these simple models.}

\section{Spherical perturbations}\label{sec:even_sector}
In this Section, we study spherical perturbations about the background metric~\eqref{eq:corrected_BG_PG} based on the simple subclass of the EFT described by the action~\eqref{eq:action_EFT} with the EFT coefficients given by Eq.~\eqref{eq:dic_scordatura}.
Note that perturbations around the stealth Schwarzschild--(anti-)de Sitter solution in DHOST theories were studied in \cite{deRham:2019gha,Khoury:2020aya,Takahashi:2021bml}, where it was found that the perturbations are strongly coupled (see also \cite{Motohashi:2019sen}).
In order to avoid this strong coupling problem, one has to take into account a small detuning of the degeneracy conditions, which can be realized by the so-called scordatura mechanism~\cite{Motohashi:2019sen,DeFelice:2022xvq}.
As mentioned earlier, the scordatura mechanism is implemented in our EFT action.

We stress that studying spherical (or monopole) perturbations is an important step to understand the dynamics of even-parity perturbations with generic multipoles.
(Note that odd- and even-parity perturbations evolve independently on a spherically symmetric background unless there is a parity-violating operator in the EFT.
The authors of \cite{Mukohyama:2022skk} studied the odd-parity perturbations based on the EFT framework and derived the master equation, i.e., the generalized Regge-Wheeler equation.)
Although we shall focus on monopole perturbations in the present paper, we hope to extend our analysis to generic multipoles in a future publication.
Therefore, for future reference, we first introduce even-parity perturbations with generic multipoles, and then restrict ourselves to monopole perturbations.

Let us introduce the even-parity metric perturbations~$\delta g_{\mu\nu} = g_{\mu\nu} - \bar{g}_{\mu\nu}$ as follows:
\begin{equation}\label{eq:pert_even_PG}
  \begin{aligned}
    \delta g_{\tau\tau} &= -\bar{g}_{\tau\tau} \sum_{\ell,m} H_{0,\ell m}(\tau, r) Y_{\ell m} (\theta, \phi) \;, \\
    \delta g_{\tau r} &= \bar{g}_{\tau r} \sum_{\ell,m} H_{1, \ell m} (\tau, r) Y_{\ell m} (\theta, \phi) \;, \\
    \delta g_{rr} &= \bar{g}_{rr} \sum_{\ell,m} H_{2, \ell m} (\tau, r) Y_{\ell m} (\theta, \phi) \;, \\
    \delta g_{\tau a} &= \sum_{\ell,m} r^2 \alpha_{\ell m}(\tau, r) \hat{\nabla}_a Y_{\ell m}(\theta, \phi) \;, \\
    \delta g_{r a} &= \sum_{\ell,m} r^2 \beta_{\ell m}(\tau, r) \hat{\nabla}_a Y_{\ell m}(\theta, \phi) \;, \\
    \delta g_{ab} &= \sum_{\ell,m} r^2\big[Q_{\ell m} (\tau, r) \gamma_{ab} Y_{\ell m} (\theta, \phi) + G_{\ell m}(\tau, r)\hat{\nabla}_a \hat{\nabla}_b Y_{\ell m} (\theta, \phi)\big] \;,
\end{aligned}  
\end{equation}
where $\hat{\nabla}_a$ denotes the covariant derivative with respect to the unit two-sphere metric~$\gamma_{ab}$ and $Y_{\ell m}(\theta,\phi)$ is the spherical harmonics.
Note that we are interested in an approximately stealth Schwarzschild background, for which the background metric~$\bar{g}_{\mu\nu}$ receives a correction characterized by $\alpha_{\rm L}$ [see Eq.~\eqref{eq:corrected_BG_PG}].
For the scalar perturbation, we have 
\begin{align}\label{eq:scalar_pert_PG}
    \delta \Phi = \sum_{\ell,m} \delta\Phi_{\ell m}(\tau, r) Y_{\ell m} (\theta, \phi) \;.
\end{align}
Under an infinitesimal coordinate transformation, $x^\mu \rightarrow x^\mu + \epsilon^\mu$, with 
\begin{align}
    \epsilon^\tau = \sum_{\ell,m} T_{\ell m}(\tau, r) Y_{\ell m}(\theta, \phi) \;, \quad \epsilon^r = \sum_{\ell, m} P_{\ell m}(\tau, r) Y_{\ell m}(\theta, \phi) \;, \quad \epsilon^a = \sum_{\ell, m} \Theta_{\ell m}(\tau, r) \hat{\nabla}^a Y_{\ell m}(\theta, \phi) \;,
\end{align}
the perturbations defined in Eqs.~(\ref{eq:pert_even_PG}) and (\ref{eq:scalar_pert_PG}) are transformed as  
    \begin{equation}\label{eq:transf_PG_generic}
    \begin{aligned}
    &H_0\to H_0 + \frac{\dot{\bar{g}}_{\tau\tau}}{\bar{g}_{\tau\tau}}T + 2\dot{T} + \frac{\bar{g}_{\tau\tau}'}{\bar{g}_{\tau\tau}}P + \frac{2\bar{g}_{\tau r}}{\bar{g}_{\tau\tau}}\dot{P} \;, \\
    &H_1\to H_1-\frac{\dot{\bar{g}}_{\tau r}}{\bar{g}_{\tau r}}T-\dot{T} - \frac{\bar{g}_{\tau\tau}}{\bar{g}_{\tau r}}T'-\frac{\bar{g}_{\tau r}'}{\bar{g}_{\tau r}}P-\frac{\bar{g}_{rr}}{\bar{g}_{\tau r}}\dot{P}-P' \;, \\
    &H_2\to H_2-\frac{\dot{\bar{g}}_{rr}}{\bar{g}_{rr}}T-\frac{2\bar{g}_{\tau r}}{\bar{g}_{rr}}T'-\frac{\bar{g}_{rr}'}{\bar{g}_{rr}}P-2P' \;, \quad 
    \alpha \to \alpha - \frac{\bar{g}_{\tau\tau}}{r^2} T - \frac{\bar{g}_{\tau r}}{r^2} P - \dot{\Theta} \;, \\
    &\beta \to \beta - \frac{\bar{g}_{\tau r}}{r^2} T - \frac{\bar{g}_{rr}}{r^2} P - \Theta' \;, \quad Q\to Q-\frac{2P}{r} \;, \quad G \to G - 2 \Theta \;, \quad \delta\Phi\to\delta\Phi-qT \;,
    \end{aligned}
    \end{equation}
where a prime denotes the derivative with respect to $r$ and a dot the derivative with respect to $\tau$.
Here, we have omitted the indices~$\ell$ and $m$ as modes with different $(\ell,m)$ evolve independently.
Due to the spherical symmetry of the background metric, one can set $m=0$ without loss of generality.
Therefore, we can conveniently work with the Legendre polynomials~$P_\ell(\cos\theta)$ instead of the spherical harmonics~$Y_{\ell m}(\theta,\phi)$. Below, $H_0$, $H_1$, $H_2$, $\alpha$, $\beta$, $Q$, $G$, and $\delta\Phi$ denote the coefficients of $P_\ell(\cos\theta)$.

From the gauge transformation (\ref{eq:transf_PG_generic}), we see that choosing $P = rQ/2$ and $\Theta = G/2$ leads to $Q\to 0$ and $G\to 0$, so that we are left with six perturbations. Also, if one further sets $T = \delta\Phi/q$, we have $\delta\Phi\to 0$, which is nothing but the unitary gauge.
It is important to note that the gauge choice, $Q = G = \delta\Phi = 0$, is a complete gauge-fixing condition, which can be imposed at the level of Lagrangian (see \cite{Motohashi:2016prk} for further discussions on this issue). 
In fact, this complete gauge-fixing condition was employed in \cite{Takahashi:2021bml} to study the dynamics of even-parity perturbations in the shift- and reflection-symmetric DHOST theories.

Having said that, it was clarified in \cite{deRham:2019gha} that the following gauge condition can be more useful than the unitary gauge~$\delta\Phi=0$ for the analysis of even modes:
\begin{align}
    \bar{\nabla}^\mu\bar{\Phi} \bar{\nabla}^\nu\bar{\Phi} \, \delta g_{\mu\nu} 
    = 0 \;,
\end{align}
which can be achieved with an appropriate choice of the gauge function~$T$.
Under this gauge condition, the scalar perturbation~$\delta\Phi$ is present, and one obtains a decoupled equation for $\delta\Phi$ in a large class of DHOST theories where the scordatura term is absent.
In our setup, the above condition can be written as
    \begin{align}\label{eq:gauge_claudia_0}
    \mathcal{G}\equiv -\bar{g}_{rr} \bar{g}_{\tau\tau} H_0 + \bar{g}_{\tau r}^2 (H_2 - 2 H_1) = 0 \;,
    \end{align}
which contains the scordatura corrections to the background.
In the presence of the scordatura term, there is no guarantee that the equation for $\delta\Phi$ is decoupled (and actually it is not decoupled in this gauge), but it would still be useful to have a decoupled equation at least at the zeroth order in $\alpha_{\rm L}$.
Note that, in the Lema{\^i}tre coordinates, the above condition can be written as $\delta g_{\tau\tau} = 0$, up to terms of ${\cal O}(\alpha_{\rm L})$.
The price to pay is that the gauge fixing is incomplete, and hence there is a residual gauge degree of freedom.
Such an incomplete gauge-fixing condition cannot be imposed at the level of Lagrangian as an independent equation may be lost.
Instead, one should impose such a gauge condition after deriving EOMs.
In order to have a better control on the residual gauge degree of freedom, we refine the gauge choice~\eqref{eq:gauge_claudia_0} as
\begin{align}\label{eq:claudia_gauge_2}
    \tilde{\mathcal{G}} \equiv  A H_0 + (1 - A)(H_2 - 2 H_1) + \alpha_{\rm L} \tilde{\mathcal{G}}_1 = 0 \;,
\end{align}
with
\begin{align}
    \tilde{\mathcal{G}}_1 \equiv [(h_2 - h_1) + A(h_0 + 2h_1 - h_2) ]H_2 - \bigg[(1-A)\dot{h}_2 + A\dot{h}_0 - 2(1-A)\dot{h}_1\bigg]\frac{\delta\Phi}{q} \;.
\end{align}
We note that $\tilde{\cal G}-{\cal G}={\cal O}(\alpha_{\rm L})$.
Here, $\tilde{\cal G}_1$ has been chosen so that the gauge transformation of $\tilde{\cal G}$ becomes of the form
\begin{align}\label{eq:gauge_transf_1}
    \tilde{\mathcal{G}} &\to \tilde{\mathcal{G}} - 2 \sqrt{1 - A}\, T' + 2 \dot{T} + {\cal O}(\alpha_{\rm L}^2)\;.
\end{align}
This implies that the gauge condition $\tilde{\mathcal{G}} = 0$ is invariant [up to terms of ${\cal O}(\alpha_{\rm L}^2)$] under a gauge transformation with
\begin{align}\label{eq:gauge_extra}
    T=T\left(\tau+\int\frac{{\rm d}r}{\sqrt{1-A}}\right) \;,
\end{align}
which corresponds to the residual gauge degree of freedom mentioned above.
As we shall see later in Subsection~\ref{sec:EOMs_mono}, the argument of $T$ in the above equation is nothing but $\rho$ in the Lema{\^i}tre coordinate system.

To summarize, when one studies the dynamics of even-parity perturbations about the approximately stealth Schwarzschild background, it would be useful to choose the gauge where $\tilde{\cal G}=0$, with $\tilde{\cal G}$ given in Eq.~\eqref{eq:claudia_gauge_2}.
As shown in \cite{deRham:2019gha}, in this gauge, one would obtain a decoupled equation for $\delta\Phi$ at least at the zeroth order in $\alpha_{\rm L}$.
However, one should note that there is a residual gauge degree of freedom given by Eq.~\eqref{eq:gauge_extra}.
On top of that, one can impose two more gauge conditions~$Q=G=0$ to completely fix the gauge functions~$P$ and $\Theta$.

The discussion so far applies to generic multipoles with $\ell\ge 2$.
For monopole ($\ell=0$) and dipole ($\ell=1$) perturbations, the situation is slightly different.
Let us briefly comment on the monopole case, and the dipole case will be discussed elsewhere.
For $\ell=0$, the perturbations~$\alpha$, $\beta$, and $G$, as well as the gauge function~$\Theta$ are absent.
In this case, on top of $\tilde{\cal G}=0$, one can impose $Q=0$ with an appropriate choice of the gauge function~$P$. 
Since the gauge fixing by $Q=0$ is complete, we will impose it at the Lagrangian level.
At this stage, the remaining perturbation variables are $H_0$, $H_1$, $H_2$, and $\delta\Phi$.
Then, the incomplete gauge-fixing condition~$\tilde{\cal G}$ will be imposed after deriving the EOMs.
In what follows, we shall discuss the dynamics of monopole perturbations under these gauge conditions.

\subsection{Equations of motion for spherical perturbations}\label{sec:EOMs_mono}

In this Subsection, we derive EOMs of the spherical (monopole) perturbations.
In each EOM, there are contributions from the scordatura term, and we shall discuss which terms are relevant at the leading order in $\alpha_{\rm L}$.
It is important to note that, for monopole perturbations, only the scalar fluctuation~$\delta\Phi$ is dynamical, whereas the rest are not propagating.

As mentioned above, we impose the gauge condition~$Q=0$ at the Lagrangian level.
From the covariant action~(\ref{eq:action_scorda}), we straightforwardly obtain the EOMs for the remaining perturbation variables~$\delta \Phi$, $H_0$, $H_1$, and $H_2$, which can be schematically written as
\begin{equation}
\begin{aligned}\label{eq:EOM_pert_Scor}
    \mathcal{E}_{\delta\Phi} &= \mathcal{E}_{\delta \Phi}^{(0)} + \alpha_{\rm L}\mathcal{E}_{\delta \Phi}^{(1)} = 0 \;, \\
    \mathcal{E}_{H_0} &= \mathcal{E}_{H_0}^{(0)} + \alpha_{\rm L} \mathcal{E}_{H_0}^{(1)} = 0 \;, \\ 
    \mathcal{E}_{H_1} &= \mathcal{E}_{H_1}^{(0)} + \alpha_{\rm L} \mathcal{E}_{H_1}^{(1)} = 0 \;, \\
    \mathcal{E}_{H_2} &= \mathcal{E}_{H_2}^{(0)} + \alpha_{\rm L} \mathcal{E}_{H_2}^{(1)} = 0\;,
\end{aligned}
\end{equation}
up to the first order in $\alpha_{\rm L}$.
Here, the superscripts~$(0)$ and $(1)$ refer to the zeroth- and the first-order terms in $\alpha_{\rm L}$.
The zeroth-order parts in $\alpha_{\rm L}$ of the above equation were already obtained in, e.g.,~\cite{deRham:2019gha,Takahashi:2021bml}.
Note that, in the EOM for $\delta\Phi$, the terms responsible for the scordatura effect, which belong to $\mathcal{E}_{\delta \Phi}^{(1)}$, are assumed to be comparable to the terms in $\mathcal{E}_{\delta \Phi}^{(0)}$.
Therefore, those terms in $\mathcal{E}_{\delta \Phi}^{(1)}$ should be included in the main part of the equation. 
For the other EOMs, we will argue that the corrections linear in $\alpha_{\rm L}$ can be safely neglected and show that in the monopole analysis the metric perturbations $H_0$, $H_1$, and $H_2$ are fixed in terms of $\delta\Phi$ by using the remaining gauge freedom and a boundary condition.

Before proceeding further, let us point out the consistency among the four EOMs in Eq.~\eqref{eq:EOM_pert_Scor}, which comes from the Noether identity for the remaining gauge degree of freedom associated with the function~$T$.
(Note that this gauge degree of freedom will be used to impose the incomplete gauge-fixing condition~$\tilde{\cal G}=0$.)
Written explicitly, we have
\begin{align}\label{eq:Noe_iden}
    \frac{\dot{\bar{g}}_{\tau\tau}}{\bar{g}_{\tau\tau}}\mathcal{E}_{H_0}-2\dot{\mathcal{E}}_{H_0}-\frac{\dot{\bar{g}}_{\tau r}}{\bar{g}_{\tau r}}\mathcal{E}_{H_1}+\dot{\mathcal{E}}_{H_1} + \left(\frac{\bar{g}_{\tau\tau}}{\bar{g}_{\tau r}}\mathcal{E}_{H_1}\right)'-\frac{\dot{\bar{g}}_{rr}}{\bar{g}_{rr}}\mathcal{E}_{H_2}+\left(\frac{2\bar{g}_{\tau r}}{\bar{g}_{rr}}\mathcal{E}_{H_2}\right)'-q\mathcal{E}_{\delta\Phi} = 0 \;.
\end{align}
We derive the above identity in Appendix~\ref{sec:Noether}. 
Note that the relation above holds at any order in $\alpha_{\rm L}$.
With the explicit expression of the EOMs in Eq.~\eqref{eq:EOM_pert_Scor}, one can verify the identity~\eqref{eq:Noe_iden} up to the first order in $\alpha_{\rm L}$ by taking into account the corrections to the background metric~$h_0$, $h_1$, and $h_2$, as well as their EOMs obtained in \cite{DeFelice:2022qaz}.

Although the EOMs in Eq.~(\ref{eq:EOM_pert_Scor}) 
were derived in the PG coordinates, it is in fact more convenient to express those EOMs in the Lema{\^i}tre coordinates~$(\tau, \rho)$ [see Eq.~(\ref{eq:bg_Lemai})] when discussing the relevance of each term. 
This is because the structure of the EOMs becomes simpler in the Lema{\^i}tre coordinate system.\footnote{Having said that, if one is interested in the dynamics of perturbations at large $r$, the PG coordinate system is more useful as the metric~\eqref{eq:bg_PG} reduces to the flat-space metric.}
In terms of the derivative operators~$(\partial_\tau,\partial_\rho)$ in the Lema{\^i}tre coordinates, the gauge transformations for $H_0$, $H_1$, $H_2$ and $\delta \Phi$ can be written as follows:
    \begin{align}
    \begin{split}
    \delta \Phi &\rightarrow \delta\Phi - q T \;, \\
    H_0 &\to H_0 + \bigg(\frac{\dot{\bar{g}}_{\tau\tau} + \partial_\rho \bar{g}_{\tau\tau} }{\bar{g}_{\tau\tau}} \bigg) T + 2(\dot{T} + \partial_\rho T) + \frac{2 \bar{g}_{\tau r}}{\bar{g}_{\tau\tau}} (\dot{P} + \partial_\rho P)   \;, \\
    H_1 &\to H_1 - \bigg(\frac{\dot{\bar{g}}_{\tau r} + \partial_\rho \bar{g}_{\tau r}}{\bar{g}_{\tau r}}\bigg) T - \dot{T} - \bigg(1 + \frac{\bar{g}_{\tau\tau}}{\bar{g}_{\tau r}\sqrt{1-A}}\bigg) \partial_\rho T - \frac{\partial_\rho \bar{g}_{\tau r}}{\bar{g}_{\tau r} } \frac{P}{\sqrt{1-A}} \\
    & \hspace{0.45cm} - \frac{\bar{g}_{rr}}{\bar{g}_{\tau r}}\dot{P} - \bigg(\frac{1}{\sqrt{1-A}} + \frac{\bar{g}_{rr}}{\bar{g}_{\tau r}}\bigg) \partial_\rho P \;, \\
    H_2 &\rightarrow H_2 - \bigg(\frac{\dot{\bar{g}}_{rr} + \partial_\rho \bar{g}_{rr}}{\bar{g}_{rr}}\bigg) T - \frac{2 \bar{g}_{\tau r}}{\bar{g}_{rr}}\frac{\partial_\rho T}{\sqrt{1-A}} - \frac{\partial_\rho \bar{g}_{rr}}{\bar{g}_{rr}}\frac{P}{\sqrt{1-A}} - \frac{2 \partial_\rho P}{\sqrt{1-A}} \;.
    \end{split}\label{eq:H_2_gauge_Lemaitre}
    \end{align}
Also, when the gauge function~$P$ is fixed, the transformation of the quantity~$\tilde{\mathcal{G}}$ [defined in Eq.~(\ref{eq:claudia_gauge_2})] reads 
\begin{align}\label{eq:gauge_transf_Lemaitre}
     \tilde{\mathcal{G}} &\to \tilde{\mathcal{G}} + 2 \dot{T} + {\cal O}(\alpha_{\rm L}^2)\;.
\end{align}
From Eq.~(\ref{eq:gauge_transf_Lemaitre}), we see that the gauge condition~$\tilde{\cal G}=0$ remains the same under the gauge transformation with $T=T(\rho)$ [see also (\ref{eq:Lemaitre_trans}) and the discussion below Eq.~\eqref{eq:gauge_transf_1}].

Moreover, for convenience, let us introduce the variables~$\delta$ and $\delta_{\rm p}$ via
 \begin{align}
    \bar{g}^{\mu\nu} \partial_\mu r \partial_\nu r &\equiv 1 - \frac{r_{\rm s}(1 + \alpha_{\rm L} \delta)}{r} \;, \label{eq:delta_def} \\
    g^{\mu\nu} \partial_\mu r \partial_\nu r &\equiv 1 - \frac{r_{\rm s}(1 + \alpha_{\rm L}\delta + \delta_{\rm p})}{r} \;,  \label{eq:delta_P_def}
    \end{align}
where $\bar{g}_{\mu\nu}$ is the background metric with the scordatura corrections and $g_{\mu\nu}$ is the metric with the (spherical) perturbations included.
Notice that $\delta$ corresponds to the scordatura correction to the Misner-Sharp mass for the background metric~$\bar{g}_{\mu\nu}$, which can be written as
\begin{align}\label{eq:delta_BG}
    \delta = \bigg(1- \frac{r_{\rm s}}{r}\bigg) \bigg[h_0 + 2h_1 + \bigg(\frac{r_{\rm s}}{r} - 1\bigg)h_2\bigg] \;.
\end{align}
Also, from Eq.~(\ref{eq:delta_P_def}), up to first order in $\alpha_{\rm L}$, we obtain
\begin{align}\label{eq:delta_P_def2}
     \delta_{\rm p} = \bigg(1- \frac{r_{\rm s}}{r}\bigg)\bigg[H_0 + 2 H_1 + \bigg(\frac{r_{\rm s}}{r} - 1\bigg) H_2\bigg] + \alpha_{\rm L}\delta_{\rm p1} \;,
\end{align}
where the correction~$\delta_{\rm p1}$ is defined by
\begin{align}
    \delta_{\rm p1} &\equiv \bigg(1- \frac{r_{\rm s}}{r}\bigg)\bigg[\bigg(1- \frac{2r_{\rm s}}{r}\bigg)(h_0 + 2h_1) - 2 \bigg(1- \frac{r_{\rm s}}{r}\bigg) h_2\bigg] (H_0 + 2H_1) \nonumber \\
    & \hspace{5mm} - \frac{r}{r_{\rm s}} \bigg(1- \frac{r_{\rm s}}{r}\bigg)^2\bigg[\frac{2r_{\rm s}}{r}(h_0 + 2h_1) + \bigg(1- \frac{2r_{\rm s}}{r}\bigg)h_2\bigg]H_2 \;.
\end{align}
Note in passing that, using the gauge transformation law~(\ref{eq:transf_PG_generic}), the gauge transformation of $\delta_{\rm p}$ can be obtained as
\begin{align}\label{eq:delta_P_transform}
    \delta_{\rm p} \rightarrow \delta_{\rm p} + \frac{P}{r} - \frac{2A P'}{1 - A} - \frac{2 \dot{P}}{\sqrt{1 - A}} + \mathcal{O}(\alpha_{\rm L}) \;,
\end{align}
where terms of ${\cal O}(\alpha_{\rm L})$ have been omitted for simplicity.

Let us now proceed with Eq.~(\ref{eq:EOM_pert_Scor}).
As mentioned earlier, we now impose the gauge condition~$\tilde{\cal G}=0$, which can be used to express $H_0$ in terms of the other perturbation variables.
Moreover, we use $\delta_{\rm p}$ as one of the independent perturbation variables in place of $H_1$.
Therefore, we are left with three variables: $\delta\Phi$, $H_2$, and $\delta_{\rm p}$.
Written in terms of these variables, the EOM of $\delta \Phi$ reads\footnote{In fact, one can straightforwardly derive Eq.~(\ref{eq:delta_P_pert}) from the covariant EOM of $\Phi$: 
\begin{align*}
    \nabla_\mu (F_{0X} \nabla^\mu \Phi) - \frac{\alpha_{\rm L}}{\Lambda_{\rm s}^2} \Box^2 \Phi = 0 \;.
\end{align*}
}
\begin{align}\label{eq:delta_P_pert}
    -\delta \ddot{\Phi} + \frac{3 \sqrt{1 - A}}{2 r} \delta\dot{\Phi} + \frac{\alpha_{\rm L} \partial^4_\rho \delta \Phi}{2 q^2 \Lambda_{\rm s}^2 \bar{F}^{(0)}_{0XX} (1-A)^2} + \alpha_{\rm L} [A_1 \partial_\rho^3 + A_2 \partial_\rho^2 + A_3 \partial_\rho ] \delta\Phi = \alpha_{\rm L} S_1 \;,
\end{align}
where $\bar{F}^{(0)}_{0XX}\equiv F_{0XX}(-q^2)$.
The coefficients~$A_1$, $A_2$ and $A_3$ are given by
\begin{equation}\label{eq:a3}
\begin{aligned}
    A_1 &= \frac{7}{2 q^2 r \Lambda_{\rm s}^2 \bar{F}^{(0)}_{0XX} (1 - A)^{3/2} }\;, \\ 
    A_2 &= \frac{19 + 4A}{8 q^2 r^2 \Lambda_{\rm s}^2 \bar{F}^{(0)}_{0XX} (1-A)} + \frac{\delta}{2(1 - A)} - \frac{h_1}{1 - A} + \frac{(1-2A)h_2}{(1-A)^2}\;, \\ 
     A_3 &= -\frac{5(2 - A)}{16 q^2 r^3 \Lambda_{\rm s}^2 \bar{F}^{(0)}_{0XX} \sqrt{1-A}} + \frac{5\delta}{4 r \sqrt{1-A}} + \frac{\partial_\rho\delta}{2(1-A)} + \bigg(\partial_\tau - \frac{\partial_\rho}{1 - A}\bigg)h_1 \\
     & \hspace{0.45cm} - \bigg[\partial_\tau - \frac{(1-2A)\partial_\rho}{2 (1-A)^2} \bigg] h_2 - \frac{(8-3A)h_1}{2r \sqrt{1-A}} + \frac{(9-22A + 6A^2)h_2}{4r (1-A)^{3/2}} \;.
\end{aligned}
\end{equation}
Here, we have replaced the correction~$h_0$ to the background metric with $\delta$, $h_1$, and $h_2$ using the formula~(\ref{eq:delta_BG}), 
and the source term~$S_1$ on the right-hand side of (\ref{eq:delta_P_pert}) has the form
\begin{align}
    S_1 &= [A_4 \partial_\tau^4 + A_5 \partial_\tau^2 \partial_\rho^2 + A_6 \partial_\tau^3 + A_7 \partial_\tau^2 \partial_\rho + A_8 \partial_\tau \partial_\rho^2 + A_{9} \partial_\tau \partial_\rho + A_{10} \partial_\tau + A_{11}] \delta\Phi \nonumber \\
    & \hspace{0.3cm} + \big[B_1 \partial_\rho^3 + B_2 \partial_\tau^2 \partial_\rho + B_3 \partial_\tau^2  + B_4 \partial_\rho^2 + B_5 \partial_\tau \partial_\rho + B_6 \partial_\tau + B_7 \partial_\rho + B_{8} \big] \delta_{\rm p} \nonumber \\ 
     & \hspace{0.3cm} + \big[C_1 \partial_\tau^3 + C_2 \partial_\rho^3 + C_3 \partial_\tau^2 \partial_\rho + C_4 \partial_\tau \partial_\rho^2 + C_5 \partial_\tau^2  + C_6 \partial_\rho^2 + C_7 \partial_\tau \partial_\rho + C_8 \partial_\tau + C_9 \partial_\rho + C_{10} \big] H_2 \;,\label{eq:S_1}
\end{align}
where the term of $\delta\ddot{\Phi}$ has been absorbed into an overall rescaling of the EOM (\ref{eq:delta_P_pert}).
We omit the explicit expressions for the coefficients~$\{A_4, \cdots, A_{11}\}$, $\{B_1,\cdots, B_{8}\}$, and $\{C_1,\cdots, C_{10}\}$ due to their length. 

Let us now comment on Eq.~(\ref{eq:delta_P_pert}). 
First, as expected, in the absence of scordatura term (i.e., $\alpha_{\rm L} = 0$), the scalar perturbation would be strongly coupled in the asymptotic flat region as the spatial-derivative terms vanish~\cite{Motohashi:2019ymr}.
In order to avoid such a problem, one needs to take into account the scordatura term, giving rise to those spatial-derivative terms, in particular the fourth-order one, on the left-hand side of (\ref{eq:delta_P_pert}).
Since the term of $\partial_\rho^2\delta\Phi$ is typically suppressed by the Planck scale, the leading spatial-derivative term in Eq.~\eqref{eq:delta_P_pert} is given by $\partial_\rho^4\delta\Phi$.
Nevertheless, in the vicinity of the BH, the lower spatial-derivative terms can in principle be non-negligible compared with the term of $\partial_\rho^4\delta\Phi$.
This is the reason why we have kept the terms of $\partial_\rho^3\delta\Phi$, $\partial_\rho^2\delta\Phi$, and $\partial_\rho\delta\Phi$ on the left-hand side of Eq.~\eqref{eq:delta_P_pert}.

After some manipulations, the EOMs for $\delta_{\rm p}$ can be written in the following form:
\begin{align}\label{eq:delta_P1}
    \dot{\delta}_{\rm p} + \alpha_{\rm L}[a_1 \partial_\rho^3 + a_2 \partial_\rho^2 + a_3 \partial_\rho] \delta\Phi = \alpha_{\rm L}S_2 \;,
\end{align}
and 
\begin{align}\label{eq:delta_P2}
    \partial_\rho \delta_{\rm p}  - \frac{4 q^3 r \bar{F}^{(0)}_{0XX} \delta\dot{\Phi}}{M_{\rm Pl}^2 \sqrt{1-A}}  + \alpha_{\rm L}[\bar{a}_1 \partial_\rho^3 + \bar{a}_2 \partial_\rho^2 + \bar{a}_3 \partial_\rho] \delta\Phi= \alpha_{\rm L} S_3 \;,
\end{align}
where the source terms $S_2$ and $S_3$ have the form
\begin{align}
    S_2 &= \big[a_4 \partial_\tau^3 + a_5 \partial_\tau \partial_\rho^2 + a_6 \partial_\tau^2 \partial_\rho + a_7 \partial_\tau^2 + a_8 \partial_\tau \partial_\rho + a_9 \partial_\tau \big] \delta\Phi + \big[b_1 \partial_\rho^2 + b_2 \partial_\tau \partial_\rho + b_3 \partial_\rho + b_4 \big] \delta_{\rm p} \nonumber \\
    & \hspace{5mm}
    + \big[c_1 \partial_\tau^2 + c_2 \partial_\rho^2 + c_3 \partial_\tau \partial_\rho + c_4 \partial_\tau + c_5 \partial_\rho + c_6 \big] H_2 \;, \label{eq:S2}
\end{align}
and 
\begin{align}
    S_3 &= \big[\bar{a}_4 \partial_\tau^3 + \bar{a}_5 \partial_\tau \partial_\rho^2 + \bar{a}_6 \partial_\tau^2 \partial_\rho + \bar{a}_7 \partial_\tau \partial_\rho + \bar{a}_8 \partial_\tau+ \bar{a}_9 \big] \delta\Phi 
    + \big[\bar{b}_1 \partial_\rho^2 + \bar{b}_2 \partial_\tau \partial_\rho + \bar{b}_3 \partial_\tau  + \bar{b}_4 \big] \delta_{\rm p} \nonumber \\
    & \hspace{5mm} + \big[\bar{c}_1 \partial_\tau^2 + \bar{c}_2 \partial_\rho^2 + \bar{c}_3 \partial_\tau \partial_\rho + \bar{c}_4 \partial_\tau + \bar{c}_5 \partial_\rho + \bar{c}_6 \big] H_2 \;. \label{eq:S3}
\end{align}
Here, we omit the explicit expressions for the coefficients in $S_2$ and $S_3$.
Note that, upon using Eq.~\eqref{eq:delta_P_pert}, one can confirm that Eqs.~\eqref{eq:delta_P1} and \eqref{eq:delta_P2} are consistent with the integrability condition~$[\partial_\tau,\partial_\rho]\delta_{\rm p}=0$.
Therefore, one can simply solve (\ref{eq:delta_P2}) for $\delta_{\rm p}$ without need for an initial condition, once $\delta\Phi$ and $H_2$ are specified and a boundary condition consistent with (\ref{eq:delta_P1}) is provided. In particular, in the absence of the scordatura term (i.e., $\alpha_{\rm L} = 0$), Eq.~\eqref{eq:delta_P1} implies that the variable~$\delta_{\rm p}$ depends only on $\rho$, and the $\rho$-dependence can be fixed via Eq.~\eqref{eq:delta_P2} under an appropriate time-independent boundary condition.

Finally, the EOM for $H_2$ can be written as
\begin{align}\label{eq:EOM_H2}
    \dot{H}_2 = \alpha_{\rm L} S_4 \;,
\end{align}
where 
\begin{align}
    S_4 &= \big[\tilde{a}_1 \partial_\rho^3 + \tilde{a}_2 \partial_\tau^2 \partial_\rho + \tilde{a}_3 \partial_\rho^2 + \tilde{a}_4 \partial_\tau \partial_\rho + \tilde{a}_5 \partial_\tau + \tilde{a}_6 \partial_\rho + \tilde{a}_7\big] \delta\Phi 
     + \big[\tilde{b}_1 \partial_\rho^2 + \tilde{b}_2 \partial_\rho + \tilde{b}_3 \big] \delta_{\rm p} \nonumber \\
    & \hspace{5mm} + \big[ \tilde{c}_1 \partial_\rho^2 + \tilde{c}_2 \partial_\tau \partial_\rho + \tilde{c}_3 \partial_\rho + \tilde{c}_4 \big] H_2 \;.
\end{align}
Here, we omit the explicit expressions for $\{\tilde{a}_1,\cdots,\tilde{a}_7\}$, $\{\tilde{b}_1, \cdots, \tilde{b}_3\}$, and $\{\tilde{c}_1, \cdots, \tilde{c}_4\}$ due to their length.
Equation~\eqref{eq:EOM_H2} fixes $H_2$ in the form of
    \begin{align}\label{eq:homo_H2}
    H_2=\alpha_{\rm L}H_2^{(1)}(\tau,\rho)+I(\rho)\;,
    \end{align}
where one can impose the condition~$H_2^{(1)}(\tau_0,\rho)=0$ at an initial time $\tau=\tau_0$ without loss of generality and $I(\rho)=H_2(\tau_0,\rho)$ is a $\rho$-dependent integration constant corresponding to the initial condition for $H_2$ at $\tau=\tau_0$.
We recall that there is a residual gauge degree of freedom~$T=T(\rho)$ due to the fact that the gauge condition~$\tilde{\cal G}=0$ is incomplete [see Eq.~\eqref{eq:gauge_transf_Lemaitre}], under which $H_2$ is transformed as
    \begin{align}\label{eq:H2_extra_gauge}
    H_2 \rightarrow H_2 - 2 \partial_\rho T + {\cal O}(\alpha_{\rm L})\;.
    \end{align}
This can be used to remove the $\rho$-dependent integration constant in Eq.~\eqref{eq:homo_H2}.
After this manipulation, the function~$H_2$ is at most of ${\cal O}(\alpha_{\rm L})$, and hence one can disregard all the contributions associated with $H_2$ in the source terms of the other EOMs [i.e., Eqs.~\eqref{eq:delta_P_pert}, \eqref{eq:delta_P1}, and \eqref{eq:delta_P2}]. This manipulation also makes it evident that an initial condition for $H_2$ is not needed for the physical description of the system.

Up to this point, we have written down the EOMs for the perturbation variables~$\delta\Phi$, $H_2$, and $\delta_{\rm p}$ up to ${\cal O}(\alpha_{\rm L})$, and shown that $H_2$ can be removed by use of the residual gauge degree of freedom.
As mentioned earlier, for spherical perturbations, one expects that only the scalar fluctuation~$\delta\Phi$ propagates.
In the next Subsection, we will obtain a decoupled EOM for $\delta\Phi$ via field redefinition, which serves as the master equation for the spherical perturbations.
Moreover, we will derive the EOMs for $\delta_{\rm p}$ sourced by $\delta\Phi$.

\subsection{Decoupled master equation}\label{sec:redef}
In this Subsection, we perform a field redefinition on the variable~$\delta\Phi$ such that its EOM~(\ref{eq:delta_P_pert}) becomes decoupled from metric perturbations.
It is important to note that, for the reason stated at the end of the previous Subsection, we shall discard the EOM for $H_2$ and remove all the contributions associated with $H_2$ from the EOMs, i.e., Eqs.~\eqref{eq:delta_P_pert}, \eqref{eq:delta_P1}, and \eqref{eq:delta_P2}, by using the residual gauge degree of freedom. 

Before proceeding further, let us simplify the source terms $S_1$, $S_2$ and $S_3$ using the zeroth-order part of Eqs.~(\ref{eq:delta_P1}) and (\ref{eq:delta_P2}).
By doing so, the source term $S_1$ [Eq.~(\ref{eq:S_1})] reduces to 
\begin{align}
  S_1 &=   [A_4 \partial_\tau^4 + A_5 \partial_\tau^2 \partial_\rho^2 + A_6 \partial_\tau^3 + A_7 \partial_\tau^2 \partial_\rho + \hat{A}_8 \partial_\tau \partial_\rho^2 + \hat{A}_{9} \partial_\tau \partial_\rho + \hat{A}_{10} \partial_\tau + A_{11}]\delta \Phi  + B_8 \delta_{\rm p} \label{eq:S_1_simplified} \;,
\end{align}
where the terms with $\dot{\delta}_{\rm p}$ and its derivatives have been removed using the zeroth-order part of (\ref{eq:delta_P1}) as they become higher order in $\alpha_{\rm L}$, and the terms with $\rho$-derivative(s) acting on $\delta_{\rm p}$ have been rewritten in terms of $\delta\Phi$ and its derivatives using the zeroth-order part of (\ref{eq:delta_P2}).
Notice that some of the coefficients in $S_1$ get modified by this manipulation, which is indicated by a hat, i.e., $\hat{A}_8\partial_\tau \partial_\rho^2 \delta \Phi$, $\hat{A}_{9}\partial_\tau\partial_\rho \delta\Phi$, and $\hat{A}_{10}\partial_\tau\delta\Phi$.
We present all of the coefficients on the right-hand side of Eq.~(\ref{eq:S_1_simplified}) in Appendix~\ref{app:S_1}.
It should be noted that the term~$B_8 \delta_{\rm p}$ cannot be removed by use of the zeroth-order EOMs of $\delta_{\rm p}$.

Let us comment on the term~$A_{11} \delta \Phi$ which does not respect the shift symmetry of the scalar field.
This term essentially comes from the fact that we used the gauge condition~\eqref{eq:claudia_gauge_2}, in which $\delta\Phi$ shows up without derivatives.
However, it is in fact straightforward to verify the coefficient~$A_{11}$ [Eq.~(\ref{app:A_coeff})] identically vanishes when imposing EOMs of the background corrections~$h_0, h_1, h_2$ (and $\delta$). Note that their EOMs in PG coordinates can be found in \cite{DeFelice:2022qaz}.
(See also footnote~\ref{footnote:BGcorrections}.)
Therefore, in what follows, we discard $A_{11}$ in $S_1$.

Following the same procedure, the source terms~$S_2$ and $S_3$ in Eqs.~\eqref{eq:delta_P1} and \eqref{eq:delta_P2} become
\begin{align}
     S_2 &= \big[a_4 \partial_\tau^3 + a_5 \partial_\tau \partial_\rho^2 + a_6 \partial_\tau^2 \partial_\rho + a_7 \partial_\tau^2 + \hat{a}_8 \partial_\tau \partial_\rho + \hat{a}_9 \partial_\tau \big] \delta\Phi + b_4 \delta_{\rm p} \;, \label{eq:S2_simplified} \\
    S_3 &= \big[\bar{a}_4 \partial_\tau^3 + \bar{a}_5 \partial_\tau \partial_\rho^2 + \bar{a}_6 \partial_\tau^2 \partial_\rho + \hat{\bar{a}}_7 \partial_\tau \partial_\rho + \hat{\bar{a}}_8 \partial_\tau+ \bar{a}_9 \big] \delta\Phi + \bar{b}_4 \delta_{\rm p} \;, \label{eq:S3_simplified}
\end{align}
where, as before, the coefficients with a hat are those that have been modified upon using the zeroth-order EOMs.\footnote{One could in principle use Eq.~\eqref{eq:delta_P_pert} to remove the term~$a_7\delta\ddot{\Phi}$ in $S_2$. However, this introduces a term of the form~$\alpha_{\rm L}^2\partial_\rho^4\delta\Phi$ in Eq.~\eqref{eq:delta_P2}. In order to avoid this complication, we keep the term with $\delta\ddot{\Phi}$.}
In Appendices~\ref{app:S_2} and \ref{app:S_3}, we present the explicit expression for the coefficients.

Let us now perform the following field redefinition of $\delta\Phi$:
\begin{align}\label{eq:redef_phi}
    \delta\Phi = \delta\tilde{\Phi} + 
    \alpha_{\rm L}(\gamma_1 \delta\ddot{\tilde{\Phi}} + \gamma_2 \delta\dot{\tilde{\Phi}} +  \gamma_3 \partial_\rho^2\delta\tilde{\Phi} + \gamma_4 \partial_\rho \delta\tilde{\Phi} +  \gamma_5 \delta_{\rm p})\;,
\end{align}
and rewrite the EOMs in terms of $\delta\tilde{\Phi}$ and $\delta_{\rm p}$.
Here, the coefficients~$\gamma_i$'s are chosen to make the EOM for $\delta\tilde{\Phi}$ as simple as possible.
Written explicitly,
\begin{equation}\label{eq:gamma_1}
\begin{aligned}
    &\gamma_1 = -A_4(\tau,\rho) \;, \quad \gamma_2 = -A_6(\tau,\rho) + 2 \dot{A}_4(\tau,\rho) - \frac{3\sqrt{1-A}}{2r}\,A_4(\tau,\rho) \;, \\
    &\gamma_3 = -A_5(\tau,\rho) \;, \quad \gamma_4 = -A_7(\tau,\rho) \;,
\end{aligned}
\end{equation}
and $\gamma_5$ satisfies the following differential equation:
\begin{align}\label{eq:gamma_5}
    -\ddot{\gamma}_5 + \frac{3\sqrt{1-A}}{2r}\, \dot{\gamma}_5 - B_8 = 0 \;.
\end{align}
Then, Eq.~(\ref{eq:delta_P_pert}) becomes
\begin{align}\label{eq:redefined_EOM_Phi}
    -\delta\ddot{\tilde{\Phi}} + \frac{3\sqrt{1-A}}{2r} \delta\dot{\tilde{\Phi}} + \frac{\alpha_{\rm L} \partial^4_\rho \delta \tilde{\Phi}}{2 q^2 \Lambda_{\rm s}^2 \bar{F}^{(0)}_{0XX} (1-A)^2} + \alpha_{\rm L} [\tilde{A}_1 \partial_\rho^3 + \tilde{A}_2 \partial_\rho^2 + \tilde{A}_3 \partial_\rho ] \delta\tilde{\Phi} = \alpha_{\rm L} \tilde{S}_1[\delta\tilde{\Phi}] \;,
\end{align}
where we have used the EOMs~\eqref{eq:delta_P1} and \eqref{eq:delta_P2} to remove the derivatives of $\delta_{\rm p}$ and kept terms of up to first order in $\alpha_{\rm L}$.
The new source term~$\tilde{S}_1$ in Eq.~(\ref{eq:redefined_EOM_Phi}) reads
\begin{align}
    \tilde{S}_1[\delta\tilde{\Phi}] &= [\tilde{A}_8 \partial_\tau \partial_\rho^2 + \tilde{A}_{9} \partial_\tau \partial_\rho + \tilde{A}_{10} \partial_\tau]\delta \tilde{\Phi} \;,
\end{align}
where the coefficients are given by 
\begin{equation}
   \begin{aligned}
    &\tilde{A}_1 = A_1 \;, \quad \tilde{A}_2 = A_2 + \ddot{A}_5 - \frac{3\sqrt{1-A}}{2r}\, \dot{A}_5 \;, \quad \tilde{A}_3 = A_3 + \ddot{A}_7 - \frac{3\sqrt{1-A}}{2r}\, \dot{A}_7 \;, \\
    &\tilde{A}_8 = \hat{A}_8 - 2 \dot{A}_5 + \frac{3\sqrt{1-A}}{2r}\, A_5 \;, \quad  \tilde{A}_{9} = \hat{A}_9 - 2 \dot{A}_7 + \frac{3\sqrt{1-A}}{2r}\, A_7 \;, \\
    &\tilde{A}_{10} = \hat{A}_{10} + 2\dddot{A}_4 - \ddot{A}_6 - \frac{3\sqrt{1-A}}{2r}\, \dot{A}_6 - \frac{9(1-A)}{r^2}\, \dot{A}_4  + \frac{9(1-A)}{4r^2}\, A_6 - \frac{27(1-A)^{3/2}}{4r^3}\, A_4 \;.
    \end{aligned}
\end{equation}

In addition to the EOM for $\delta\tilde{\Phi}$, Eqs.~(\ref{eq:delta_P1}) and (\ref{eq:delta_P2}) for $\delta_{\rm p}$ become
\begin{align}
    \dot{\delta}_{\rm p} - \alpha_{\rm L} b_4 \delta_{\rm p} &= \alpha_{\rm L}\tilde{S}_2[\delta\tilde{\Phi}] \;, \label{eq:simplified_delta_P1} \\
    \partial_\rho \delta_{\rm p} - \alpha_{\rm L} q_4 \delta_{\rm p} &= + \frac{4 q^3 r \bar{F}^{(0)}_{0XX} \delta\dot{\tilde{\Phi}}}{M_{\rm Pl}^2 \sqrt{1-A}} + \alpha_{\rm L} \tilde{S}_3[\delta\tilde{\Phi}] \;, \label{eq:simplified_delta_P2}
\end{align}
respectively, where the new source terms $\tilde{S}_2$ and $\tilde{S}_3$ are given by
\begin{align}
     \tilde{S}_2[\delta\tilde{\Phi}] &= \big[-a_1 \partial_\rho^3 - a_2 \partial_\rho^2 - a_3 \partial_\rho + a_4 \partial_\tau^3 + a_5 \partial_\tau \partial_\rho^2 + a_6 \partial_\tau^2 \partial_\rho + a_7 \partial_\tau^2 + \hat{a}_8 \partial_\tau \partial_\rho + \hat{a}_9 \partial_\tau \big] \delta\tilde{\Phi} \;, \label{eq:redef_S_2} \\
    \tilde{S}_3[\delta\tilde{\Phi}] &= \big[\,p_0\partial_\tau^2-p_1 \partial_\rho^3 - p_2 \partial_\rho^2 - p_3 \partial_\rho + p_4 \partial_\tau^3 + p_5 \partial_\tau \partial_\rho^2 + p_6 \partial_\tau^2 \partial_\rho + p_7 \partial_\tau \partial_\rho + p_8 \partial_\tau + p_9 \big] \delta\tilde{\Phi} \;. \label{eq:redef_S_3}
\end{align}
The coefficients in $\tilde{S}_2$ remain the same as those given in Eq.~(\ref{app:eq:S2_1}). 
The coefficients introduced in Eqs.~(\ref{eq:simplified_delta_P2}) and \eqref{eq:redef_S_3} are given by
\begin{equation}
\begin{aligned}
    &q_4 = \bar{b}_4 + \frac{4q^3 r \bar{F}^{(0)}_{0XX} \dot{\gamma}_5}{M_{\rm Pl}^2 \sqrt{1-A}} \;, \quad 
    p_0 = \frac{4q^3 r \bar{F}^{(0)}_{0XX} (\gamma_2-\dot{A}_4)}{M_{\rm Pl}^2 \sqrt{1-A}}\;, \quad
    p_1 = \bar{a}_1 \;, \quad p_2 = \bar{a}_2 + \frac{4q^3 r \bar{F}^{(0)}_{0XX} \dot{A}_5}{M_{\rm Pl}^2 \sqrt{1-A}} \;, \\
    &p_3 = \bar{a}_3 + \frac{4q^3 r \bar{F}^{(0)}_{0XX} \dot{A}_7}{M_{\rm Pl}^2 \sqrt{1-A}} \;, \quad
    p_4 = \bar{a}_4 - \frac{4q^3 r \bar{F}^{(0)}_{0XX} A_4}{M_{\rm Pl}^2 \sqrt{1-A}} \;, \quad
    p_5 = \bar{a}_5 - \frac{4q^3 r \bar{F}^{(0)}_{0XX} A_5}{M_{\rm Pl}^2 \sqrt{1-A}} \;, \quad
    p_6 = \bar{a}_6 \;, \\
    &p_7 = \hat{\bar{a}}_7 - \frac{4q^3 r \bar{F}^{(0)}_{0XX} A_7}{M_{\rm Pl}^2 \sqrt{1-A}} \;, \quad
    p_8 = \hat{\bar{a}}_8 + \frac{4q^3 r \bar{F}^{(0)}_{0XX} \dot{\gamma}_2}{M_{\rm Pl}^2 \sqrt{1-A}} \;, \quad
    p_9 = \bar{a}_9 \;.
\end{aligned}
\end{equation}
Similarly to Eqs.~\eqref{eq:delta_P1} and \eqref{eq:delta_P2}, Eqs.~\eqref{eq:simplified_delta_P1} and \eqref{eq:simplified_delta_P2} are consistent with the integrability condition~$[\partial_\tau,\partial_\rho]\delta_{\rm p}=0$. One can therefore solve \eqref{eq:simplified_delta_P2} for $\delta_{\rm p}$ without need for an initial condition, once $\delta\tilde{\Phi}$ is specified and a boundary condition consistent with \eqref{eq:simplified_delta_P1} is provided.

Finally, let us discuss the structure of the system of EOMs~\eqref{eq:redefined_EOM_Phi}, \eqref{eq:simplified_delta_P1}, and \eqref{eq:simplified_delta_P2}.
As mentioned earlier, Eq.~\eqref{eq:redefined_EOM_Phi} is a decoupled equation for $\delta\tilde{\Phi}$, which can in principle be solved by itself.
Then, Eqs.~\eqref{eq:simplified_delta_P1} and \eqref{eq:simplified_delta_P2} fix the profile of $\delta_{\rm p}$ in terms of $\delta\tilde{\Phi}$ under an appropriate boundary condition without need for an extra initial condition.
Moreover, it is straightforward to reconstruct all the original perturbation variables up to gauge transformations.
In this sense, Eq.~\eqref{eq:redefined_EOM_Phi} serves as the master equation of spherical perturbations about the (approximately) stealth Schwarzschild solution.
Having said that, it is still non-trivial how to study, e.g., the ringdown waveform due to the presence of a higher spatial-derivative term.\footnote{See \cite{Oshita:2021onq} for a time-domain analysis of the ringdown waveform of a Lifshitz scalar field, whose EOM involves up to the sixth-order spatial derivatives.}
We leave this issue for future work.

\section{Conclusions}\label{sec:conclusion}
We have analyzed perturbations about the (approximately) stealth Schwarzschild background based on the effective field theory (EFT) of black hole perturbations with a timelike scalar profile~\cite{Mukohyama:2022enj}.
As pointed out in \cite{Motohashi:2019ymr}, perturbations about stealth solutions in DHOST theories are plagued by the strong coupling problem in general, and one has to take into account the so-called scordatura term, which corresponds to a small detuning of (one of) the degeneracy conditions.
The EFT framework of \cite{Mukohyama:2022enj} is useful in this context as the scordatura term is involved in the action by default.
For concreteness, as explained in Section~\ref{sec:setup}, we have considered a simple subclass of the EFT that accommodates the k-essence model with the scordatura term [see Eq.~\eqref{eq:action_scorda}], i.e.~ghost condensation.
The scordatura effect is controlled by a dimensionless parameter~$\alpha_{\rm L}$.
In the presence of the scordatura term, the stealth Schwarzschild configuration cannot be realized as an exact solution, but one can obtain an approximately stealth solution perturbatively with respect to $\alpha_{\rm L}$~\cite{DeFelice:2022qaz}.
As expected, the scordatura term does not qualitatively affect the dynamics of odd-parity perturbations, and it is rather straightforward to derive the master equation based on the EFT framework~\cite{Mukohyama:2022skk}.
On the other hand, the analysis for even-parity perturbations is still lacking due to complications coming from the scordatura term.
In order to pave the way to understand the dynamics of the even modes, in Section~\ref{sec:even_sector}, we have investigated spherical (monopole) perturbations on the approximately stealth Schwarzschild background.
We have derived the EOMs for the perturbations up to the linear order in $\alpha_{\rm L}$, keeping higher spatial derivatives responsible for the scordatura effect.
More importantly, with an appropriate choice of gauge conditions and field redefinition~$\delta\Phi\to\delta\tilde{\Phi}$, we have shown that the EOM for $\delta\tilde{\Phi}$ can be decoupled, which is the master equation that governs the dynamics of the spherical perturbations.
The metric perturbations, which are non-dynamical for the spherical perturbations, are fixed in terms of $\delta\tilde{\Phi}$ under an appropriate boundary condition.

There are several future directions we would like to explore.
First, as already mentioned in Section~\ref{sec:even_sector}, it is important to extend our analysis to even-parity perturbations with generic higher multipoles.
In this case, on top of the EOM for the scalar perturbation, we would obtain a generalized Zerilli equation that describes the dynamics of metric perturbations.\footnote{In general, the two dynamical degrees of freedom in the even sector propagate with different speeds, which would lead to interesting phenomena such as energy
extraction from the black hole and a characteristic late-time relaxation~\cite{Cardoso:2024qie}.}
We expect that our analysis for spherical perturbations is an essential step in this direction.
It would also be intriguing to extend our analysis to a more generic background metric.
Second, of course after obtaining the generalized Zerilli equation, it would be interesting to investigate, e.g., the quasinormal modes and the ringdown waveform, to see how the result is deviated from that of GR.
In particular, we expect that the isospectrality between the odd and even sectors no longer holds in scalar-tensor theories due to the presence of the scalar degree of freedom.
Lastly, as an extension of the analysis for the odd modes in \cite{Barura:2024uog}, it would also worth exploring the tidal response of BHs (characterized by the tidal Love numbers and the dissipation numbers) for the even modes.
We leave these issues for future studies.

\section*{Acknowledgements}

This work was supported by World Premier International Research Center Initiative (WPI Initiative), MEXT, Japan.
S.M.~was supported in part by JSPS KAKENHI Grant No.~JP24K07017. 
K.Ta.~was supported in part by JSPS KAKENHI Grant No.~JP23K13101.
K.To.~was supported in part by the National Key Research and Development Program of China under Grant No.~2020YFC2201504.
V.Y.~is supported by grants for development of new faculty staff, Ratchadaphiseksomphot Fund, Chulalongkorn University.


\appendix

\section{Noether identity for the action of spherical perturbations}\label{sec:Noether}
In this Appendix, we explicitly derive the Noether identity for the action of spherically symmetric (or monopole) perturbations.
As pointed out in Section~\ref{sec:EOMs_mono}, the Noether identity serves as a consistency check among the EOMs for the perturbations. 

To derive the identity, let us recall that the approximately stealth Schwarzschild background~\cite{DeFelice:2022qaz}\footnote{Note that in \cite{DeFelice:2022qaz} the authors used $\{H_0, H_1, H_2\}$ for spherical corrections to the background, but here we use $\{h_0, h_1, h_2\}$.} takes the following form, up to the first order in $\alpha_{\rm L}$ [see Eq.~(\ref{eq:corrected_BG_PG})]:
    \begin{equation} \label{eq:corrected_BG_app}
    \bar{g}_{\tau\tau}= -A(1-\alpha_{\rm L}h_0)\;, \qquad
    \bar{g}_{\tau r}=\sqrt{1-A}\,(1+\alpha_{\rm L}h_1)\;, \qquad
    \bar{g}_{rr}=1+\alpha_{\rm L}h_2\;, \qquad
    \bar{g}_{ab}=r^2\gamma_{ab} \;.
    \end{equation}
Here, $A = 1- r_{\rm s}/r$ with $r_{\rm s}\,(>0)$ being a constant and the deviation from the Schwarzschild metric is controlled by the constant parameter~$\alpha_{\rm L}$, which is associated with the scordatura effect.
Note that $h_0$, $h_1$, and $h_2$ are functions of $(\tau,r)$.

We then introduce the monopole perturbations ($H_0$, $H_1$, $H_2$, $Q$, and $\delta\Phi$) about the above background, as in Eqs.~(\ref{eq:pert_even_PG}) and (\ref{eq:scalar_pert_PG}).
Note that, for monopole perturbations, the variables~$\alpha$, $\beta$, and $G$ are intrinsically absent.
Furthermore, under an infinitesimal coordinate transformation~$x^\mu \rightarrow x^\mu + \epsilon^\mu$, the perturbations are transformed as in Eq.~(\ref{eq:transf_PG_generic}).
For monopole perturbations, the gauge function~$\Theta$ is irrelevant, and one can fix $Q=0$ by use of the gauge function~$P$.
For the reason explained in the main text, we do not fix the gauge function~$T$ from the outset, and we shall discuss the Noether identity associated with it.

With the setup above, the quadratic action for monopole perturbations can be schematically written in the PG coordinates as    
    \begin{equation}
    S^{(2)}_{\rm even}
    =\int {\rm d}\tau{\rm d}r\,\mathcal{L}^{(2)}_{\rm even}[H_0,H_1,H_2,\delta\Phi] \;.
    \label{S_even}
    \end{equation}
This action must be invariant under the remaining gauge transformation, i.e.,
   \begin{equation}
    \begin{aligned}\label{eq:trans_pert_app}
    &H_0\to H_0 +  \frac{\dot{\bar{g}}_{\tau\tau}}{\bar{g}_{\tau\tau}}T + 2\dot{T} \;, \qquad
    H_1\to H_1-\frac{\dot{\bar{g}}_{\tau r}}{\bar{g}_{\tau r}}T-\dot{T} - \frac{\bar{g}_{\tau\tau}}{\bar{g}_{\tau r}}T' \;,  \\
    &H_2\to H_2-\frac{\dot{\bar{g}}_{rr}}{\bar{g}_{rr}}T-\frac{2\bar{g}_{\tau r}}{\bar{g}_{rr}}T' \;, \qquad 
    \delta\Phi\to\delta\Phi-qT \;.
    \end{aligned}
    \end{equation}
The change of the action~\eqref{S_even} under the above remaining gauge transformation can be written as
    \begin{align}
    \delta S^{(2)}_{\rm even}
    =\int {\rm d}\tau{\rm d}r\bigg[\!\left(\frac{\dot{\bar{g}}_{\tau\tau}}{\bar{g}_{\tau\tau}}T + 
    2\dot{T}\right)\mathcal{E}_{H_0} &+ \left(-\frac{\dot{\bar{g}}_{\tau r}}{\bar{g}_{\tau r}}T-\dot{T} -  \frac{\bar{g}_{\tau\tau}}{\bar{g}_{\tau r}}T'\right)\mathcal{E}_{H_1} \nonumber \\
    &+\left(-\frac{\dot{\bar{g}}_{rr}}{\bar{g}_{rr}}T-\frac{2\bar{g}_{\tau r}}{\bar{g}_{rr}}T'\right)\mathcal{E}_{H_2} -qT\mathcal{E}_{\delta\Phi}\bigg] \;,
    \label{deltaS_even}
    \end{align}
where $\mathcal{E}_{\cal Q}$ denotes the left-hand side of the Euler-Lagrange equation associated with ${\cal Q}\in\{H_0,H_1,H_2,\delta\Phi\}$.
After integration by parts, Eq.~\eqref{deltaS_even} can be rewritten as
    \begin{align}
    \delta S^{(2)}_{\rm even}
    =\int {\rm d}\tau{\rm d}r\bigg[\frac{\dot{\bar{g}}_{\tau\tau}}{\bar{g}_{\tau\tau}}\mathcal{E}_{H_0}-2\dot{\mathcal{E}}_{H_0}-\frac{\dot{\bar{g}}_{\tau r}}{\bar{g}_{\tau r}}\mathcal{E}_{H_1}+\dot{\mathcal{E}}_{H_1} &+ \left(\frac{\bar{g}_{\tau\tau}}{\bar{g}_{\tau r}}\mathcal{E}_{H_1}\right)'
    -\frac{\dot{\bar{g}}_{rr}}{\bar{g}_{rr}}\mathcal{E}_{H_2} \nonumber \\ 
    &+\left(\frac{2\bar{g}_{\tau r}}{\bar{g}_{rr}}\mathcal{E}_{H_2}\right)'-q\mathcal{E}_{\delta\Phi}\bigg]T \;.
    \end{align}
Since this should vanish for any choice of the function~$T$, we find that the following combination among $\mathcal{E}_{\cal Q}$'s should vanish identically:
    \begin{equation}\label{eq:Noether_id}
   \mathcal{I}_{\rm N} = \frac{\dot{\bar{g}}_{\tau\tau}}{\bar{g}_{\tau\tau}}\mathcal{E}_{H_0}-2\dot{\mathcal{E}}_{H_0}-\frac{\dot{\bar{g}}_{\tau r}}{\bar{g}_{\tau r}}\mathcal{E}_{H_1}+\dot{\mathcal{E}}_{H_1} + \left(\frac{\bar{g}_{\tau\tau}}{\bar{g}_{\tau r}}\mathcal{E}_{H_1}\right)'-\frac{\dot{\bar{g}}_{rr}}{\bar{g}_{rr}}\mathcal{E}_{H_2}+\left(\frac{2\bar{g}_{\tau r}}{\bar{g}_{rr}}\mathcal{E}_{H_2}\right)'-q\mathcal{E}_{\delta\Phi} = 0 \;.
    \end{equation}

One can straightforwardly check that, at the zeroth-order in $\alpha_{\rm L}$, the identity~(\ref{eq:Noether_id}) is automatically satisfied. 
At the first order in $\alpha_{\rm L}$, on the other hand, the check of the identity~(\ref{eq:Noether_id}) is more involved, as one needs to use the EOMs for the corrections $h_0$, $h_1$, and $h_2$, derived in \cite{DeFelice:2022qaz}. 
More explicitly, after plugging the EOMs defined in Eq.~(\ref{eq:EOM_pert_Scor}) into Eq.~(\ref{eq:Noether_id}), the left-hand side of the Noether identity~$\mathcal{I}_N$ at $\mathcal{O}(\alpha_{\rm L})$ can be schematically recast into the following form: 
\begin{align}\label{eq:noether_first}
    \mathcal{I}_{\rm N}^{(1)} = {\cal A}_1 H_0 + {\cal A}_2 H_1 + {\cal A}_3 \dot{H}_0 + {\cal A}_4 H'_0 + {\cal A}_5 H'_1 + {\cal A}_6 \dot{H}_2 + {\cal A}_7 \delta \dot{\Phi} \;,
\end{align}
where the coefficients ${\cal A}_1, {\cal A}_2, \cdots, {\cal A}_7$ involve the corrections to the background (i.e., $h_0$, $h_1$, and $h_2$).
Then, we have confirmed that all the seven coefficients vanish upon using the equations for $h_0$, $h_1$, and $h_2$.

\section{Coefficients in the source terms of the equations of motion}\label{app:coeff}
\subsection{\texorpdfstring{$S_1$}{S1}}\label{app:S_1}
\label{app:negativezeta}
Here, we express all the coefficients which appear in the source term~$S_1$ in Eq.~(\ref{eq:S_1_simplified}).
We have 
\begin{equation}\label{app:A_coeff}
\begin{aligned}
    A_4 &= -\frac{1}{2}(1-A)A_5 = -\frac{rA_6}{3\sqrt{1-A}}
    = - \frac{1}{5} r A_7 \sqrt{1-A} = - \frac{1}{2 q^2  \Lambda_{\rm s}^2 \bar{F}^{(0)}_{0XX}} \;, \\
    \hat{A}_8 &= - \frac{5(1-A) + 2 \xi_2}{2 q^2 r \Lambda_{\rm s}^2 \bar{F}^{(0)}_{0XX} (1-A)^{3/2}} \;, \qquad 
    \hat{A}_{9} = - \frac{1 - A + 28\xi_2}{4 q^2 r^2 \Lambda_{\rm s}^2 \bar{F}^{(0)}_{0XX}(1-A)} - 2 (h_1 - h_2) \;, \\
    \hat{A}_{10} &= \frac{3[15 - 40\xi_2 -3(11 - 4\xi_2)A + 18 A^2]}{16 q^2 r^3 \Lambda_{\rm s}^2 \bar{F}^{(0)}_{0XX}\sqrt{1-A}} - \frac{(3 - 4\xi_2 - 3A)h_1}{2r\sqrt{1-A}} - \frac{\xi_2 \delta}{r \sqrt{1-A}}\\ 
     & \hspace{5mm} - \frac{[7-4\xi_1 - 2\xi_2 - 2(5 - 2\xi_1 - 2 \xi_2)A + 3A^2]h_2}{2r (1-A)^{3/2}}  + \bigg(\frac{5}{2} -2 \xi_1 \bigg) \dot{\delta} \\ 
    & \hspace{5mm}
    + \frac{\partial_\rho \delta}{2}+ \frac{[6 - 4\xi_1 - (11-8\xi_1)A]\dot{h}_2}{2(1-A)} + \frac{(3-4A)\partial_\rho h_2}{2(1-A)} - \big[2\partial_\rho + (5 - 4\xi_1 )\partial_\tau \big] h_1 \;, \\
    A_{11} &= \frac{\ddot{\delta}}{2} - \ddot{h}_1 + \frac{(1 - 2 A )\ddot{h}_2}{2(1-A)} + \frac{\partial_\rho \dot{\delta}}{2} - \partial_\rho \dot{h}_1    + \frac{(1-2A)\partial_\rho \dot{h}_2}{2(1-A)} - \frac{3\sqrt{1-A}}{4r}(\partial_\rho + \partial_\tau) \delta   \\
    & \hspace{5mm}  + \frac{3}{2r} \sqrt{1-A}(\partial_\rho + \partial_\tau) h_1 - \frac{(1-6A)}{4r\sqrt{1-A}}(\partial_\rho + \partial_\tau) h_2 \;, \\
     B_8 &= \frac{9(8-9A)\sqrt{1-A}}{32q r^3 \Lambda_{\rm s}^2 \bar{F}^{(0)}_{0XX}} - \frac{q(4 - 10A + 9A^2)\sqrt{1-A}\,\delta}{8 r A^2}  + \frac{3q \sqrt{1-A}\,h_1}{4r} - \frac{q(7-12A)h_2}{8r\sqrt{1-A}} \\
    & \hspace{5mm} - \frac{q}{4} \bigg[\partial_\rho + \frac{2(1-A)}{A}\,\partial_\tau\bigg] \delta + \frac{1}{2}\partial_\rho h_1
    + \frac{q}{2} \bigg[\partial_\tau - \frac{1 - 2A}{2(1-A)}\,\partial_\rho\bigg] h_2 \;,
\end{aligned}
\end{equation}
where $\xi_1 \equiv q^2 \bar{F}^{(0)}_{0XXX}/(2\bar{F}^{(0)}_{0XX})$ and $\xi_2(r) \equiv q^4 r^2 \bar{F}^{(0)}_{0XX}/M_{\rm Pl}^2$. 
Note that the symbols~$\bar{F}^{(0)}_{0XX}$ and $\bar{F}^{(0)}_{0XXX}$ refer to the functions~$F_{0XX}$ and $F_{0XXX}$ evaluated at $X=-q^2$.

\subsection{\texorpdfstring{$S_2$}{S2}}\label{app:S_2}
Here, we display all the coefficients in the source term~$S_2$ of Eq.~(\ref{eq:S2_simplified}). 
We have
\begin{equation}\label{app:eq:S2_1}
    \begin{aligned}
        a_1 &= - \frac{2qr}{M_{\rm Pl}^2 \Lambda_{\rm s}^2 (1-A)^{5/2}} \;, \\
        a_3 &= \frac{5q(3-2A)}{2r M_{\rm Pl}^2 \Lambda_{\rm s}^2 (1-A)^{3/2}} - \frac{2\xi_2}{qr(1-A)^{3/2}}\bigg[\delta - 2h_1 + \frac{(1-2A)h_2}{1-A}\bigg] \;, \\
        \hat{a}_8 &=  \frac{4 q (2 + \xi_2 - 2 A)}{M_{\rm Pl}^2 \Lambda_{\rm s}^2(1-A)^2} \;, \quad 
        \hat{a}_9 = -\frac{3q(1+A)(3-4\xi_2-3A)}{2r M_{\rm Pl}^2 \Lambda_{\rm s}^2 (1-A)^{3/2}} + \frac{2 \xi_2 \delta}{qr\sqrt{1-A}} - \frac{2 \xi_2 h_2}{qr(1-A)^{3/2}}  \;, \\
        b_4 &= -\frac{9q^2(1+A)}{8r M_{\rm Pl}^2 \Lambda_{\rm s}^2 \sqrt{1-A}} + \frac{\xi_2\delta}{r\sqrt{1-A}} - \frac{2\xi_2 h_1}{r \sqrt{1-A}} + \frac{[2+ 2\xi_2 - A(1+4\xi_2)]h_2}{2r(1-A)^{3/2}} \\ 
       & \hspace{5mm} + \frac{1}{2}\bigg( \partial_\tau - \frac{2-A}{1-A}\,\partial_\rho \bigg)\delta + \frac{(2-A)\partial_\rho h_1}{1-A}  - \frac{1}{2(1-A)} \bigg[(2-A)\partial_\tau + \frac{2 - 5A + 2A^2}{1-A}\,\partial_\rho\bigg] h_2 \;,
    \end{aligned}
\end{equation}
with the following relations:
\begin{equation}\label{app:eq:S2_2}
        a_1 = \frac{2r a_2}{(2+5A)\sqrt{1-A}} = \frac{a_4}{(1-A)^2} = - \frac{a_5}{1-A} = \frac{a_6}{1-A} = -\frac{ra_7}{3(1-A)^{5/2}} \;.
\end{equation}

\subsection{\texorpdfstring{$S_3$}{S3}}\label{app:S_3}
Here, we express all the coefficients in the source term~$S_3$ of Eq.~(\ref{eq:S3_simplified}).
We have 
\begin{equation}
    \begin{aligned}
        \bar{a}_1 &= \frac{r\bar{a}_2}{4\sqrt{1-A}} = \frac{\bar{a}_4}{1-A} = - \bar{a}_5 = \frac{\bar{a}_6}{1-A}  = -\frac{2 q r}{M^2_{\rm Pl}\Lambda_{\rm s}^2 (1-A)^{3/2}} \;, \\
        \bar{a}_3 &= -\frac{3q(5-3A)}{2r M^2_{\rm Pl}\Lambda_{\rm s}^2\sqrt{1-A}} - \frac{2\xi_2}{qr\sqrt{1-A}} \bigg[\delta - 4 h_1 + \frac{(3-4A)h_2}{1-A}\bigg] \;, \\
        \hat{\bar{a}}_7 &= \frac{q(8 + 4\xi_2 - 3A)}{M_{\rm Pl}^2 \Lambda_{\rm s}^2(1-A)}\;, \\
        \hat{\bar{a}}_8 &= \frac{9q(1+4\xi_2 - A)}{2r M_{\rm Pl}^2 \Lambda_{\rm s}^2\sqrt{1-A}} - \frac{2(5-4\xi_1)\xi_2 (2h_1 - \delta)}{qr \sqrt{1-A}} + \frac{2(5-4\xi_1)\xi_2(1-2A)h_2}{qr(1-A)^{3/2}} \;, \\ 
   \bar{a}_9 &= \frac{2 \xi_2 (\partial_\rho + \partial_\tau)\delta}{qr \sqrt{1-A}} - \frac{4\xi_2 (\partial_\rho + \partial_\tau)h_1}{qr \sqrt{1-A}} + \frac{2\xi_2 (1-2A) (\partial_\rho + \partial_\tau)h_2}{qr (1-A)^{3/2}} \;, \\
   \bar{b}_4 &= \frac{9q^2(7-3A)\sqrt{1-A}}{8r M^2_{\rm Pl}\Lambda_{\rm s}^2} - \frac{\xi_2(2-A)\sqrt{1-A}\,\delta}{rA} - \frac{2\xi_2 h_1 \sqrt{1-A}}{r} - (2-A)\partial_\rho h_1 \\
        & \hspace{5mm} - \frac{[2 - 6\xi_2 - (1 - 4\xi_2 )A]h_2}{2r \sqrt{1-A}} + \frac{1}{2}\partial_\rho\delta  + \frac{1}{2}\bigg[(2-A)\partial_\tau + \frac{2-5A+2A^2}{1-A}\,\partial_\rho\bigg] h_2 \;.
    \end{aligned}
\end{equation}

\bibliographystyle{utphys}
\bibliography{bib_v4}

\end{document}